\documentclass[dvipsnames, twocolumn]{aastex631}

\usepackage{xcolor}
\usepackage{comment}
\usepackage{orcidlink}

\begin{document}

\title{A deep search for Complex Organic Molecules toward the protoplanetary disk of V883 Ori}

\author[0009-0003-6626-8122]{Abubakar M. A. Fadul}
\affiliation{Max Planck Institute for Astronomy, Königstuhl 17, D-69117 Heidelberg, Germany}

\author[0000-0002-6429-9457]{Kamber R. Schwarz}
\affiliation{Max Planck Institute for Astronomy, Königstuhl 17, D-69117 Heidelberg, Germany}

\author[0000-0002-2555-9869]{Merel L. R. van ’t Hoff}
 \affiliation{Department of Physics and Astronomy, Purdue University, 525 Northwestern Ave, West Lafayette, IN 47907, USA}

 \author[0000-0001-6947-6072]{Jane Huang}
 \affiliation{Department of Astronomy, Columbia University, 538 W. 120th Street, Pupin Hall, New York, NY 10027, USA}

 \author[0000-0002-8716-0482]{Jennifer B. Bergner}
 \affiliation{UC Berkeley Department of Chemistry, Berkeley, CA 94720, USA}

\author[0000-0002-4755-4719]{Tushar Suhasaria}
\affiliation{Max Planck Institute for Astronomy, Königstuhl 17, D-69117 Heidelberg, Germany}

 \author[0000-0002-0150-0125]{Jenny K. Calahan}
 \affiliation{University of Michigan, 323 West Hall, 1085 South University Ave., Ann Arbor, MI 48109, USA}
 \affiliation{Center for Astrophysics | Harvard \& Smithsonian, 60 Garden St., Cambridge, MA 02138, USA}








\begin{abstract}
Complex Organic Molecules (COMs) in the form of prebiotic molecules are potentially building blocks of life. Using Atacama Large Millimeter/submillimeter Array (ALMA) Band 7 observations in spectral scanning mode, we carried out a deep search for COMs within the disk of V883 Ori, covering frequency ranges of $\sim$ 348 -- 366 GHz. V883 Ori is an FUor object currently undergoing an accretion burst, which increases its luminosity and consequently increases the temperature of the surrounding protoplanetary disk, facilitating the detection of COMs in the gas phase. We identified 26 molecules, including 14 COMs and 12 other molecules, with first detection in this source of the molecules: $\mathrm{CH_3OD}$, $\mathrm{H_2 C^{17}O}$, and $\mathrm{H_2 ^{13}CO}$. We searched for multiple nitrogen-bearing COMs, as $\mathrm{CH_3CN}$ had been the only nitrogen-bearing COM that has been identified so far in this source. We also detected $\mathrm{CH_3CN}$, and tentatively detect CH$_3$CH$_2$CN, $\mathrm{CH_2CHCN}$, $\mathrm{CH_3OCN}$, $\mathrm{CH_3NCO}$, and $\mathrm{NH_2CHO}$. We compared the abundances relative to $\mathrm{CH_3OH}$ with those in the handful of objects with previous detections of these species: the Class 0 protostars IRAS 16293-2422 A, IRAS 16293-2422 B and B1-c, the high-mass star-forming region Sagittarius B2 (North), the Solar System comet 67P/Churyumov-Gerasimenko, and the protoplanetary disk of Oph-IRS 48. We report $\sim$ 1 to 3 orders of magnitude higher abundances compared to Class 0 protostars and $\sim$ 1 to 3 orders of magnitude lower abundances compared to the protoplanetary disk, Sagittarius B2 (North), and 67P/C-G. These results indicate that the protoplanetary disk phase could contribute to build up of COMs.


\end{abstract}

\keywords{Astrochemistry --- Protoplanetary discs --- FU Orions --- Complex Organic Molecules (COMs) }


\section{Introduction} \label{sec:intro}

COMs are molecules that have at least six atoms, including at least one carbon atom \citep{herbst2009ARA&A}. The identification of these species is critical for understanding the chemical evolution of star-forming regions and investigating the connection between the initial phases of star formation and the formation of solar system objects, as well as understanding the chemistry that influences the formation of planets that occurs after the disk formation around young stellar objects (YSOs) (\citealt{Calcutt2018A&A, Ruizrodr2022MNRAS.515.2646R}). Some of the key astrochemistry questions are: under which  conditions can COMs exist, which prebiotic molecules, a subclass of complex organic molecules that are thought to be the building blocks of life, can survive during the formation and evolution of stars and planetary objects, and how? (\citealt{Ritson2012NatCh, Powner2009, Henning2013ChRv, Schwarz2018ApJ}) To answer such questions, it is important to study and understand the chemical evolution of protoplanetary disks, as they are the birthplace of planets. The formation of COMs is thought to occur mainly via reactions on dust grain surfaces. Additionally, heavier radicals with multiple atoms can form through UV/X-ray/cosmic-ray-induced photolysis and hydrogen abstraction of large molecules. These radicals then diffuse and react with one another, leading to the formation of more complex molecules, via radical-radical reactions (\citealt{garrod2006A&A, herbst2009ARA&A,Chuang2016, Shingledecker2018ApJ}). These reactions are already expected to occur in molecular clouds and protostellar envelopes.

Thanks to the ALMA and the James Webb Space Telescope (JWST) with their high sensitivity, spectral, and angular resolution, we are able to reveal and advance our understanding of the chemical complexity and planet formation within disks (\citealt{Perotti2024, Schwarz2021ApJS, Lee2019ApJa, Nashanty2022A&A, Bergner2018ApJ, Ilee_2021}). COMs have been detected in the interstellar medium (ISM) around high- and low-mass protostars in the gas phase in hot and dense regions (\citealt{Charnley1992, Turner1991}, \citealt{Blake1987ApJ, Cazaux_2003, herbst2009ARA&A, Belloche2013A&A}). In hot corinos, COMs are found in the warm inner envelopes surrounding the protostars, rather than in a fully formed protoplanetary disk, as they are in a more embedded phase where the surrounding cloud material plays a key role in molecule formation and desorption processes (\citealt{Charnley1992, Turner1991}, \citealt{Cazaux_2003, Jrgensen2016A&A, Jorgensen2020ARA&A, McGuire2022ApJS, Hsu2022ApJ}). Recent studies have detected numerous COMs around these types of sources (e.g., \citealt{Calcutt2018A&A, Jorgensen2018A&A, Jrgensen2016A&A, Martin-Dom2021ApJ, Chen2024}). COMs have also been observed in protoplanetary disks, which represent later evolutionary stages. Disks such as those around Oph-IRS 48, V883 Ori, HD 100546, and HD 169142 exhibit a rich chemical complexity in the gas phase (\citealt{Lee2019ApJa, Lee_2019, Marel2021A&A, Booth2021A&A, Nashanty2022A&A, Leemker2023A&A, yamato2023chemistry, Booth2021NatAs, Booth2023A&A, Lee2025arXiv}, \citealt{Oberg2021}). These observations have significantly advanced our understanding of the chemistry in star-forming regions and, in particular, the processes occurring within protoplanetary disks.

V883 Ori is an FU Orionis object located in the Orion A L1641 molecular cloud in transition between Class I and Class II with a thin envelope and a massive disc (\citealt{Alarcon2024}). It is currently undergoing a rapid increase in luminosity due to a burst of accretion onto the central star. The mass of the central star was estimated to be 1.3 $M_{\odot}$ embedded with a well-developed Keplerian rotating disk of mass $\gtrsim$ 0.3 $M_{\odot}$ and a distance of 388 pc, as well as a bolometric luminosity of 200 $L_{\odot}$ (\citealt{Greene_2008, Cieza2016Natur.535..258C, Furlan_2016, Lee_2019, Ruizrodr2022MNRAS.515.2646R}). The high luminosity of the central protostar increases the disk temperature, which liberates the molecules that were frozen out on the dust grain surfaces. This allows us to detect these molecules in the gas phase. This mechanism is known as thermal desorption (\citealt{vantHoff2018, tobin2023deuterium}). This makes sources such as V883 Ori an excellent target for the detection of freshly sublimated COMs, as recent studies have shown that (\citealt{Lee_2019, yamato2023chemistry, Lee2025arXiv, tobin2023deuterium,vantHoff2018, Jenny2024}). \citet{Cieza2016Natur.535..258C} suggest that the water snow line at the disk midplane of V883 Ori is located within 42 au from the central star based on the intensity break seen at the $0.1\arcsec$ continuum emission. The snow line may extend up to $\sim$160 au on the disk surface (\citealt{Lee_2019, Leemker2021A&A, vantHoff2018}). However, recent observations of HDO by \cite{tobin2023deuterium} found that the water snow line at the midplane exists at a radius of 80 au, while on the disk surface, it extends up to 160 au.

 Recent studies have detected numerous COMs within the disk of V883 Ori, with the first detection of $\mathrm{CH_3OH}$ in this disk reported by \cite{vantHoff2018}. \cite{Lee_2019} identified five COMs in this disk, while \cite{yamato2023chemistry} reported the detection of 11 COMs (including isotopologues). However, \citet{Lee_2019, yamato2023chemistry, Lee2025arXiv} showed that nitrogen-bearing COMs (N-bearing COMs) are deficient in this source, with only the detection of $\mathrm{CH_3CN}$. In this work, we present a deep search for additional and more complex COMs, and N-bearing COMs within the protoplanetary disk of V883 Ori. N-bearing complex organic molecules (COMs) are of particular interest, as nitrogen plays a key role in the development of life \citep{Powner2009}. 
 
 The structure of this work is as follows: In Section \ref{sec:reduction}, we give an overview of the observation and data reduction processes, followed by the methodology that we utilize in order to identify different molecules, presented in Section \ref{sec:analysis}. Afterwards, we present our findings in Section \ref{sec:results}. Furthermore, we discuss the results that we have obtained in Section \ref{discussion}. Finally, we draw a conclusion of our work in Section \ref{sec:conclusion}.

\section{Observations and Data Reduction} \label{sec:reduction}
The observations of V883 Ori were carried out with ALMA in Band 7 over six separate execution blocks (EBs) between December 30, 2021, and September 19, 2022,  as part of ALMA Cycle 8 (project code: 2021.1.00452.S, PI: Kamber Schwarz). The observation spanned the frequency range of 348 -- 366 GHz in spectral scan mode, covering a large number of COM transitions. More information about the observations, such as the observation dates, precipitable water vapor (PWV), baseline coverage, number of antennas used, and integration time on-source, can be found in Table \ref{tab:observations}. The correlator configuration for each execution block included four spectral windows (SPWs), each with a bandwidth of 938 MHz. The central frequencies of the SPWs were tuned to 348.468, 349.375, 350.281, and 351.187 GHz. The spectral windows were configured with a channel width of 488.281 kHz, providing a velocity resolution of approximately 0.42 km $\mathrm{s^{-1}}$. The phase center was at $(\alpha\, , \delta)_{J2000} \,=\, (05^{h}28^{m}18.1023^{s} \, , -007^{\circ}02\arcmin 26.010\arcsec)$.

The data were processed using the standard ALMA calibration pipeline version 2022.2.0.64. Imaging was performed using the Common Astronomy Software Applications (CASA) version 6.4.1.12 \citep{The_CASA_Team_2022}, without applying self-calibration. The quasar J0423-0120 was used as the flux and bandpass calibrator for the first two execution blocks (EBs), while J0538-4405 was used for the remaining four EBs. For phase calibration, the quasar J0541-0541 was used in three EBs, and J0607-0834 was used in the other three. Final imaging was performed using the \textit{tclean} task with \textit{Briggs} weighting. A \textit{Robust} factor of 0.5 was used for the visibility weighting, which allows for a balance between spatial resolution and sensitivity. A Gaussian fit was performed on the continuum image using \textit{imfit} task. The resulting beam size for the continuum image is $0.46\arcsec \times 0.27\arcsec$ with a position angle of $-70.0^{\circ}$ and an RMS of 1.55 mJy beam$^{-1}$. The resulting beam size for the line cube is $0.48\arcsec \times 0.29\arcsec$ with a position angle of $-68^{\circ}$ and an RMS of 5.97 mJy beam$^{-1}$. The line-free channels were selected by identifying frequency ranges within the spectral windows that showed no significant spectral lines or emission features. The continuum subtraction was performed automatically in the image space by the ALMA pipeline, and the frequencies are listed in the CASA log file.

\begin{deluxetable*}{cccccccc}[!htbp]
\tablecaption{Observational Details\label{tab:observations}}
\tablewidth{0pt}
\tablehead{
\colhead{Obs. Date} & 
\colhead{PWV} & 
\colhead{Baseline} & 
\colhead{No. of Ant.} & 
\colhead{Int. Time (On-source)} \\ 
\colhead{} & 
\colhead{(mm)} & 
\colhead{(m)} & 
\colhead{} & 
\colhead{(min)} } 

\startdata
2021-12-30 & 0.8 & 14.9--783.1 & 42 & 38 \\
2022-01-06 & 0.6 & 14.9--976.6 & 44 & 37.98 \\
2022-04-2 & 0.8 & 14.9--500.2 & 45 & 38 \\
2022-05-06 & 0.7 & 15.1--500.2 & 43 & 37.92 \\
2022-09-03 & 0.5 & 15.3--783.5 & 39 & 37.97 \\
2022-09-19 & 0.5 & 15.1--500.2 & 42 & 37.98 \\
\enddata
\end{deluxetable*}

\section{Analysis} \label{sec:analysis}

\begin{center}
    \begin{figure*}[!htbp]
    \gridline{\fig{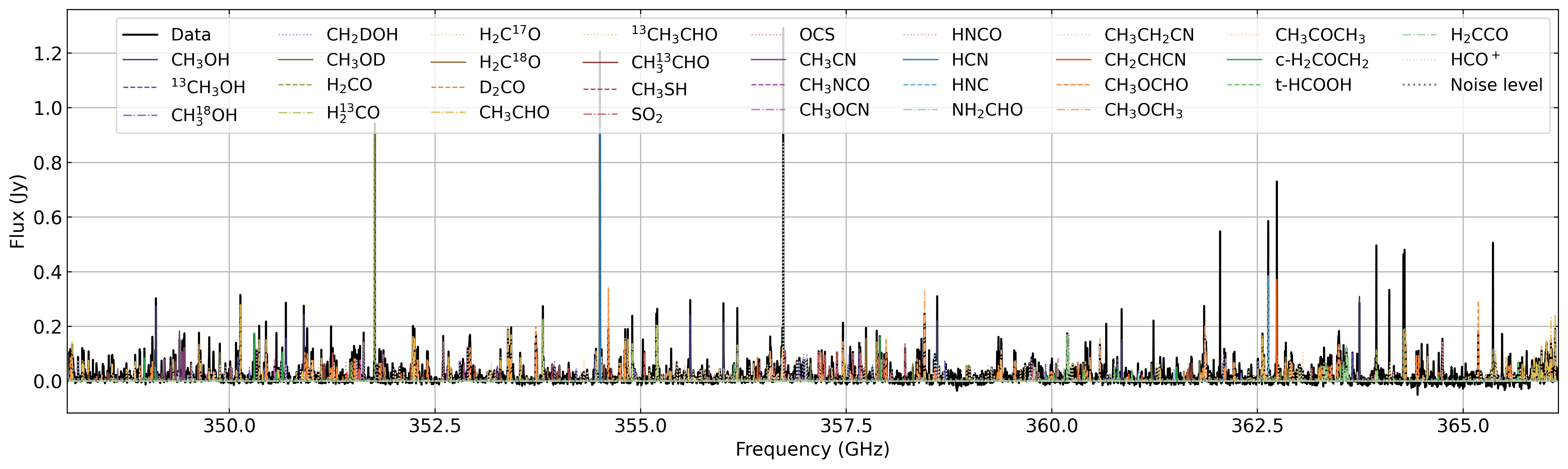}{1.0\textwidth}{}}
    \vspace{-20pt}
    \caption{Identified COMs towards the disk of V883 Ori. The horizontal dotted line in gray represents the noise level in the data ($\mathrm{1\sigma}$), the black line represents the data, and different colors represent the best fit of the model of different species.\label{analysis}}
    \end{figure*}
\end{center}

 Figure \ref{analysis} shows the full spectrum of V883 Ori after continuum subtracted, with all detected molecules indicated. The detailed spectra can be found in Appendix \ref{AppendixA}, and the accumulative model summing each molecule's contribution can be found in Appendix \ref{AppendixB}. We obtained the spectra from the image cube by using the function \textit{integrated\_spectrum()} implemented in the Python package Gofish \citep{GoFish}. This function extracts and integrates spectral information across a specified spatial region within the image cube and returns the spectra in units of Jansky (Jy). The function deprojects the disk to correct for the geometric distortions and then correct for the velocity shift due to the Keplerian rotation of the disk, recovering a single peak from a double-peaked spectrum. Consequently, this decreases the overlap of transitions and facilitates the identification of blended transitions. The main parameters we used to extract the spectra are: an inner radius of 0\arcsec, an outer radius of 0.6\arcsec, a position angle of $32^{\circ}$, an inclination angle of $38.3^{\circ}$, a distance of 388 pc, and a stellar mass of 1.29 $M_\odot$ (\citealt{Cieza2016Natur.535..258C, Lee_2019, Tobin2023, yamato2023chemistry}). To identify the different species of the COMs, we fitted synthetic spectrum to the observed spectrum. The molecular data are obtained from the Cologne Database for Molecular Spectroscopy (CDMS; \citealt{Endres_2016, 2001A&A...370L..49M}) and the Jet Population Laboratory database (JPL; \cite{1998JQSRT..60..883P}), using ``Splatalogue'' online database\footnote{https://splatalogue.online/\#/home}.

\subsection{Model}
\indent To generate the synthetic spectra we use a model with the following free parameters: excitation temperature (T$_{ex}$), the emitting radius (R$_{\mathrm{emit}}$), the distance to the star (d), and the total column density (N$_{\mathrm{tot}}$) to fit the lines assuming Local Thermodynamic Equilibrium (LTE) conditions. However, we do not assume that the lines are optically thin, and we include the optically thin and optically thick transitions in the model. We estimated the noise level in the data ($\sigma$) by calculating rms in free line emission area in the image cube using CASA. We obtained the total model flux for each molecule in Janskies based on the formula 

\begin{equation}
    F_{mod}(\nu) = \Omega \, \Delta V \, B(\nu, T_{ex}) \, \frac{\nu \, \tau_0}{2 \, c \, \sqrt{\text{ln}(2)}},
\end{equation}

\noindent where the solid angle $\Omega$ is calculated by first determining the emission area in square arcseconds, based on the angular emission radius R and distance to the star d. This area is then converted to steradians, yielding $\Omega = \pi\, \Biggl(\frac{R}{d}\Biggl)^2$, $\Delta V $ is the full width at half maximum (FWHM) of the line, with a fixed value for all species ($\Delta V = 2$ $\mathrm{km\; s^{-1}}$) as demonstrated by \cite{Lee_2019}. This value provided a good fit for all molecular species. To evaluate the impact of fixing the FWHM on the derived parameters, we tested this assumption by both varying and fixing the FWHM for multiple molecules. The results showed no significant change in the best-fit values for $\mathrm{N_{tot}}$ or $\mathrm{T_{ex}}$. This suggests that the assumption of a fixed linewidth does not significantly affect the derived parameters in our analysis. $B(\nu, T_{ex})$ is the Planck function for the blackbody radiation, $\nu$ is the transition frequency, $c$ is the speed of light, and $\tau$ is the optical depth at the line center and it is calculated as 

\begin{equation}
    \tau_0 = \Biggl(\frac{\sqrt{\mathbf{ln(2)}}}{4\pi\, \sqrt{\pi}}\Biggl)\, \Biggl(\frac{A_{ul}\, N_{tot}}{\Delta V}\Biggl)\, \Biggl(\frac{c}{\nu}\Biggl)^3\, \Biggl(\frac{n_l\, g_u}{g_l} - n_u\Biggl),
\end{equation}

\noindent where $A_{ul}$ is the Einstein A Coefficient for spontaneous emission, $n_u$ and $n_l$ are the upper and lower states of the level population, respectively, and $g_u$ and $g_l$ are their upper and lower statistical weights.

\subsection{Fitting and Identification of COMs and simple molecules}

\indent The spectrum of each molecule is then fitted using the Markov Chain Monte Carlo (MCMC) package implemented in \textit{Emcee}  \citep{emcee2013}, which explores a range of $N_{tot}$ and $T_{ex}$. We combined the spectra from different spectral windows into a single file to fit each molecule. We used 100 walkers for 10000 steps for all species, we discarded the first 2000 steps as burn-in, the remaining 8000 are used to demonstrate the posterior distribution of the parameter space of the $T_{ex}$ and $N_{tot}$ (see appendix \ref{AppendixC}). In the fitting process, we fit one molecule at a time, identified its best fit, and then subtracted the best fit model before fitting the next molecule. The fitting and subtraction were performed in the following order: $\mathrm{CH_3OH}$, $\mathrm{^{13}CH_3OH}$, $\mathrm{CH_2DOH}$, $\mathrm{SO_2}$, $\mathrm{CH_3OCH_3}$, $\mathrm{\text{c-}H_2COCH}_2$, $\mathrm{CH_3CHO}$, $\mathrm{CH_3OCHO}$, $\mathrm{\text{t-}HCOOH}$, $\mathrm{H_2 ^{13}CO}$, $\mathrm{CH_3COCH_3}$, $\mathrm{CH_3CN}$, $\mathrm{H_2CCO}$, $\mathrm{CH_3OD}$, $\mathrm{H_2CO}$, $\mathrm{D_2CO}$, $\mathrm{HCN}$, $\mathrm{OCS}$, $\mathrm{HCO^+}$, $\mathrm{H_2C^{18}O}$, $\mathrm{HNCO}$, $\mathrm{CH_3SH}$, $\mathrm{NH_2CHO}$, $\mathrm{CH_3CH_2CN}$, $\mathrm{CH_3 ^{18}OH}$, $\mathrm{CH_2CHCN}$, $\mathrm{^{13}CH_3CHO}$, $\mathrm{CH_3 ^{13}CHO}$, $\mathrm{CH_3OCN}$, $\mathrm{CH_3NCO}$, $\mathrm{H_2C^{17}O}$, and $\mathrm{HNC}$. After obtaining the spectra for all molecules, we subtracted them from the data and refitted the first molecule ($\mathrm{CH_3OH}$) to ensure that there is no contribution from other molecules. The best fit model is determined by $\chi^2$ fit as follows

\begin{equation}\label{chi2}
    \chi^2 = -\frac{1}{2}\sum_i\frac{(F_{obs,i} - F_{mod,i})^2}{\sigma^2}, 
\end{equation}

\noindent where  $F_{obs}$ and $F_{mod}$ represent the observed and modeled line intensity respectively, in a given channel $i$, and $\sigma$ represent the rms. We constrained $\chi^2$ only in the frequencies where there is emission from a given molecule based on the initial model. Additionally, the $\chi^2$ is calculated only for observed spectra with signal levels exceeding five times the noise level (5$\sigma$). We only fit species where multiple transitions are detected. At our spatial resolution, most emission is unresolved. However, we know from previous higher spatial resolution observations \citep{Lee_2019} that the emitting area for these molecules is typically $0.3\arcsec$. Hence, all the COMs in our model are confined within an emission radius of $0.3\arcsec$ ($\sim 120$ au). The identification of a molecular species is based on the synthetic spectra, which is a simulated representation of the expected molecular transitions. For COMs, we consider a molecule to be detected if it has multiple transitions that match the data.

\subsection{Model limitation}

Line blending is a significant problem, and it becomes more pronounced with increasing data complexity. This occurs because the frequency spaces between different molecular transitions are smaller than the line broadening. As we can see from the best-fit model in Appendix \ref{AppendixA}, many lines overlap with each other. We choose to fit a single molecule at a time. Alternatively, all lines can be fitted simultaneously to effectively account for line overlap \citep{Grant_2024}. However, this approach significantly increases computational time, and the large number of lines in our data makes this method impractical. Additionally, when we subtract the fit of any poorly matched molecule from the data, residual spectra from these molecules may remain. Consequently, when fitting a new molecule, some of this residual flux may be mistakenly attributed to the new molecule. 

Furthermore, we also tried fitting only optically thin transitions for some molecules to test how the fit and the column density would change, but no significant changes were observed, neither in column density nor in the fit. For example, for the molecule HNCO, when we exclude transitions with an optical depth $> 0.5$, the column density changes from $2.57 \times 10^{15}\; \mathrm{cm^{-2}}$ to $2.70 \times 10^{15} \; \mathrm{cm^{-2}}$, and no observable change in the best-fit spectra was noticed, while for $\mathrm{CH_3CH_2CN}$ no change neither in column density nor in the fit as its transitions are optically thin ($\sim$ 0.04). Therefore, we included all optical depths to estimate the column density. Moreover, we also noticed that some molecules have detected transitions with high $\mathrm{E_u}$ ($>$ 400 K). The high-$\mathrm{E_u}$ transitions primarily trace the hotter regions, while the low-$\mathrm{E_u}$ ones trace the cooler gas. The LTE fit might then be biased toward the temperature that best fits the more prominent transitions, which may contribute to the over- and/or under-prediction of some transitions. These issues can be addressed by introducing a more complex model, such as a non-LTE model, which we leave for future work.

\begin{table*}[!htbp]
\centering
\rotatebox{90}{%
\begin{minipage}{\textheight}
\caption{The identified COMs, their best fit values, and abundances}\label{resultstable}
\begin{tabular}{lcccccc}
\hline \hline 
Molecule & \begin{tabular}{c} 
Formula \\
\end{tabular} & \begin{tabular}{c} 
$N_{tot}$ \\
$\left(\mathrm{log_{10}(cm}^{-2})\right)$
\end{tabular} & \begin{tabular}{c} 
$T_{ex}$ \\
$\left(\mathrm{K}\right)$
\end{tabular} & \begin{tabular}{c} 
$\mathrm{Abundance^\star}$ \\
X(w.r.t.  $\mathrm{CH}_3\mathrm{OH}$)
\end{tabular} & \begin{tabular}{c} 
$\mathrm{Abundance}^\dag$ \\
X(w.r.t $\mathrm{H}_2$)
\end{tabular}& \begin{tabular}{c} 
No. lines
\end{tabular} \\
\hline 
Methanol & $\mathrm{CH}_3 \mathrm{OH}$ & $ 16.48^{+0.01}_{-0.01}$ & $100.07^{+0.36}_{-0.36}$ & $-$ & $2.14 \times 10^{-9}$ & 13\\
& $\mathrm{CH_3OD}$ & $15.80^{+0.30}_{-0.25}$ & $105.93^{+5.60}_{-4.62}$ & $8.29 \times 10^{-3}$ & $4.48 \times 10^{-10}$ & 22\\
& $\mathrm{CH_2DOH}$ & $16.12^{+0.03}_{-0.03}$ & $112.19^{+4.01}_{-3.68}$ & $1.73 \times 10^{-2}$ & $9.36 \times 10^{-10}$ & 41\\
& $\mathrm{^{13}CH}_3 \mathrm{OH}$ & $ 16.10^{+0.01}_{-0.01}$ & $106.44^{+1.21}_{-1.19}$ & $1.67 \times 10^{-2}$ & $9.00 \times 10^{-10}$ & 11\\
& $\mathrm{CH_3 ^{18}OH}$ & $15.89^{+0.65}_{-0.64}$ & $[100.0]$ & $1.02 \times 10^{-2}$ & $5.51 \times 10^{-10}$ & 5\\
Acetaldehyde & $\mathrm{CH_3CHO}$ & $15.96^{+0.05}_{-0.05}$ & $328.49^{+2.46}_{-2.42}$ & $1.19 \times 10^{-2}$ & $6.44 \times 10^{-10}$ & 95\\
& $\mathrm{^{13}CH_3CHO}$ & $15.54^{+0.11}_{-0.11}$ & $[328.0]$ & $4.59 \times 10^{-3}$ & $2.48 \times 10^{-10}$ & 28\\
& $\mathrm{CH_3 ^{13}CHO}$ & $15.37^{+0.15}_{-0.15}$ & $[328.0]$ & $3.09 \times 10^{-3}$ & $1.67 \times 10^{-10}$ & 11\\
Methyl formate & $\mathrm{CH_3OCHO}$ & $16.73^{+0.05}_{-0.05}$ & $66.17^{+0.20}_{-0.20}$ & $7.07\times 10^{-2}$ & $3.82 \times 10^{-9}$ & 87\\
Dimethyl ether & $\mathrm{CH_3OCH}_3$ & $16.31^{+0.02}_{-0.02}$ & $78.34^{+0.42}_{-0.42}$ & $2.67\times 10^{-2}$ & $1.44 \times 10^{-9}$ & 13\\
Acetone & $\mathrm{CH_3COCH_3}$ & $16.24^{+0.03}_{-0.03}$ & $102.10^{+2.26}_{-2.15}$ & $2.17\times 10^{-2}$ & $1.24 \times 10^{-9}$ & 93\\
Ethylene oxide & $\mathrm{\text{c-}H_2COCH}_2$ & $15.15^{+0.01}_{-0.01}$ & $50.60^{+1.04}_{-1.00}$ & $1.85 \times 10^{-3}$ & $1.00 \times 10^{-10}$ & 8\\
Formic Acid & $\mathrm{\text{t-}HCOOH}$ & $15.13^{+0.04}_{-0.03}$ & $118.71^{+23.67}_{-17.37}$ & $1.76 \times 10^{-3}$ & $9.50 \times 10^{-11}$ & 10\\
Ketene & $\mathrm{H_2CCO}$ & $15.41^{+0.05}_{-0.05}$ & $146.62^{+4.28}_{-4.05}$ & $3.51 \times 10^{-3}$ & $1.89 \times 10^{-10}$ & 5\\
Formaldehyde & $\mathrm{H_2CO}$ & $15.22^{+0.04}_{-0.04}$ & $[138.0]$ & $2.17 \times 10^{-3}$ & $1.17 \times 10^{-10}$ & 7\\
& $\mathrm{H_2 ^{13}CO}$ & $14.94^{+0.09}_{-0.09}$ & $138.21^{+2.36}_{-2.28}$ & $1.16 \times 10^{-3}$ & $6.26 \times 10^{-11}$ & 5\\
& $\mathrm{H_2C^{17}O}$ & $14.72^{+0.27}_{-0.25}$ & $123.24^{+8.71}_{-7.74}$ & $6.87 \times 10^{-4}$ & $3.71 \times 10^{-11}$ & 3\\
& $\mathrm{H_2C^{18}O^{\ddag}}$ & $14.75^{+0.05}_{-0.05}$ & $[138.0]$ & $7.50 \times 10^{-4}$ & $4.05 \times 10^{-11}$ & 1\\
& $\mathrm{D_2CO}$ & $14.99^{+0.01}_{-0.01}$ & $[138.0]$ & $1.31\times 10^{-3}$ & $7.07 \times 10^{-11}$ & 5\\
Formylium & $\mathrm{HCO^+}$ & $14.19^{+0.01}_{-0.01}$ & $[100.0]$ & $2.06 \times 10^{-4}$ & $1.11 \times 10^{-11}$ & 1\\
Formamide & $\mathrm{NH_2CHO^{\ddag}}$ & $14.78^{+0.17}_{-0.17}$ & $[100.0]$ & $7.94 \times 10^{-4}$ & $4.29 \times 10^{-11}$ & 1\\
Ethyl Cyanide & $\mathrm{CH_3CH_2CN}$ & $15.40^{+0.03}_{-0.03}$ & $[300.0]$ & $6.87 \times 10^{-3}$ & $3.71 \times 10^{-10}$ & 9\\
Vinyl Cyanide & $\mathrm{CH_2CHCN^\ddag}$ & $16.61^{+0.03}_{-0.03}$ & $50.73^{+0.17}_{-0.18}$ & $5.34 \times 10^{-2}$ & $2.89 \times 10^{-9}$ & 5\\
Methyl isocyanate & $\mathrm{CH_3NCO^{\ddag}}$ & $16.37^{+0.09}_{-0.09}$ & $[300]$ & $1.92 \times 10^{-2}$ & $1.69 \times 10^{-9}$ & 10\\
Methyl cyanate & $\mathrm{CH_3OCN^{\ddag}}$ & $15.22^{+0.04}_{-0.04}$ & $[300]$ & $2.22 \times 10^{-3}$ & $1.20 \times 10^{-10}$ & 6\\
Acetonitrile & $\mathrm{CH_3CN}$ & $14.61^{+0.19}_{-0.19}$ & $252.30^{+9.20}_{-8.70}$ & $5.38 \times 10^{-4}$ & $2.91 \times 10^{-11}$ & 9\\
Isocyanic Acid & $\mathrm{HNCO}$ & $15.41^{+0.15}_{-0.13}$ & $311.18^{+16.89}_{-15.42}$ & $3.37 \times 10^{-3}$ & $1.82 \times 10^{-10}$ & 7\\
Hydrogen Cyanide & $\mathrm{HCN^{\ddag}}$ & $14.16^{+0.01}_{-0.02}$ & $[100.0]$ & $1.93 \times 10^{-4}$ & $1.04 \times 10^{-11}$ & 1\\
Hydrogen Isocyanide & $\mathrm{HNC^{\ddag}}$ & $14.03^{+0.01}_{-0.01}$ & $[100.0]$ & $1.43 \times 10^{-4}$ & $7.71 \times 10^{-12}$ & 1\\
Methyl Mercaptan & $\mathrm{CH_3SH}$ & $15.62^{+0.07}_{-0.07}$ & $85.49^{+4.05}_{-3.74}$ & $5.56 \times 10^{-3}$ & $3.00 \times 10^{-10}$ & 13\\
Sulfur dioxide & $\mathrm{SO}_2$ & $15.55^{+0.03}_{-0.03}$ & $93.26^{+1.19}_{-1.18}$ & $4.70 \times 10^{-3}$ & $2.54 \times 10^{-10}$ & 20\\
Carbonyl Sulfide & $\mathrm{OCS}$ & $15.68^{+1.04}_{-0.72}$ & $130.60^{+28.74}_{-20.00}$ & $6.53 \times 10^{-3}$ & $3.53 \times 10^{-10}$ & 2\\
\hline
\end{tabular}\\
$^{\dag}$ The $\mathrm{H}_2$ column density is assumed to be $1.4 \times 10^{25}$ \citep{Lee_2019}. $^\ddag$ Based on a tentative detection. $^\star$ Derived based on the $\mathrm{^{13}CH}3 \mathrm{OH}$ column density multiplied by the $\mathrm{^{12}C/^{13}C}$ ratio of 60 \citep{Langer93ApJ}. The square brackets indicate molecules for which T$_{ex}$ is not well constrained by the model, and hence their T$_{ex}$ values were fixed.
\end{minipage}%
}
\end{table*}


\section{Results} \label{sec:results}
We robustly identified 14 COMs including isotopologues: $\mathrm{CH_3OH}$, $\mathrm{^{13}CH_3OH}$, $\mathrm{CH_2DOH}$, $\mathrm{CH_3OD}$, $\mathrm{CH_3 ^{18}OH}$, $\mathrm{CH_3OCH_3}$, $\mathrm{\text{c-}H_2COCH}_2$,  $\mathrm{CH_3OCHO}$, $\mathrm{CH_3SH}$, $\mathrm{CH_3CHO}$, $\mathrm{^{13}CH_3CHO}$, $\mathrm{CH_3 ^{13}CHO}$, $\mathrm{CH_3COCH_3}$ and $\mathrm{CH_3CN}$. We also identified 12 other molecules, including isotopologues: $\mathrm{SO_2}$, $\mathrm{H_2CO}$, $\mathrm{H_2 ^{13}CO}$, $\mathrm{H_2C^{17}O}$, $\mathrm{D_2CO}$, $\mathrm{HNCO}$, $\mathrm{HCO^+}$, $\mathrm{H_2CCO}$, $\mathrm{OCS}$, $\mathrm{HCN}$, $\mathrm{HNC}$, and $\mathrm{\text{t-}HCOOH}$. Most of these species had already been identified in previous studies of this source (e.g.;\ \citealt{Lee_2019, yamato2023chemistry, vantHoff2018, lee2024ApJ, Lee2025arXiv}), except for: $\mathrm{CH_3OD}$, $\mathrm{H_2 ^{13}CO}$, and $\mathrm{H_2C^{17}O}$, which are newly identified in this study. We detected only one transition of $\mathrm{HCO^+}$, HCN and HNC, which are among the strongest in our data, and no other lines are expected in the observed frequency range. Therefore we consider these robust detection. $\mathrm{CH_3SH}$ and $\mathrm{\text{c-}H}_2\mathrm{COCH}_2$ were tentatively detected by \cite{Lee_2019}, and we robustly confirmed the detection of these molecules. Additionally, \cite{vantHoff2018} tentatively detected $\mathrm{D_2CO}$, and we robustly confirm the detection of this molecule. We detected only one transition of $\mathrm{NH_2CHO}$ and $\mathrm{H_2C^{18}O}$, therefore, they are considered tentative detection. For $\mathrm{CH_3NCO}$, $\mathrm{CH_3OCN}$, CH$_3$CH$_2$CN, and $\mathrm{CH_2CHCN}$, the majority of their transitions are overestimated by the model, while others are blended with other transition of other molecules; therefore, these are considered tentative detections.

\subsection{Excitation temperature and column density derivation}
In contrast with previous studies, which assumed a single temperature for all species, we have fit for the $T_{ex}$ of each species. For several species, the temperature is not well constrained. These include $\mathrm{NH_2CHO}$, HCN, $\mathrm{H_2CO}$, and $\mathrm{H_2C^{18}O}$, among others (see Table \ref{resultstable}). For such molecules, we fix the $T_{ex}$ to that of one of its isotopologues where the $T_{ex}$ is well constrained by the model. If none of its isotopologues are detected, then we fix $T_{ex}$ to a temperature which appears to fit the lines well. For the less abundant species such as $\mathrm{H_2C^{18}O}$ and $\mathrm{NH_2CHO}$ the difficulty in constraining $T_{ex}$ is likely because we detect relatively few lines. For the more abundant species such as $\mathrm{H_2CO}$ it is possible that our assumption of a single population in LTE is not valid. \citet{lee2024ApJ} note that several species in their higher spatial resolution Band 6 have multiple origins, including the disk, shocks, and infalling material. In particular, $\mathrm{H_2CO}$ and HCN emission are attributed to both disk and infall emission. Indeed, extended emission consistent with infall is seen in $\mathrm{HCO^+}$, $\mathrm{HCN}$, and $\mathrm{H_2CO}$ in this dataset. This emission will be explored in subsequent work. 

The column density of the best fit for each molecule, including its uncertainty is presented in Table \ref{resultstable}. The median value of the posterior is taken as the best estimate of the parameter, while the difference between the median and the 16th/84th percentiles provides the lower and upper uncertainties, respectively. The column densities span between $10^{14} - 10^{16}\; \mathrm{cm^{-2}}$, and since we include both optically thin and optically thick transitions, the column densities of some molecules are underestimated. For example, $\mathrm{CH_3OH}$ and $\mathrm{H_2CO}$ have only a few detected transitions. If we exclude the optically thick transitions, there will not be enough transitions to accurately infer the column densities of these molecules. Hence, to derive the molecular abundances, we adopted the $\mathrm{^{13}CH_3OH}$ column density multiplied by the $\mathrm{^{12}C/^{13}C}$ elemental abundance of the canonical ISM (60, \cite{Langer93ApJ}), since $\mathrm{^{13}CH_3OH}$ transitions are optically thin.

\subsection{Formaldehyde and its isotopologues}
$\mathrm{H}_2\mathrm{CO}$ has been reported in the disk of V883 Ori in previous Band 6 observations (\citealt{lee2024ApJ, Ruizrodr2022MNRAS.515.2646R}), as well as in this study. In addition, we reported three of its isotopologues: $\mathrm{H_2 ^{13}CO}$, $\mathrm{H_2C^{18}O}$, and $\mathrm{D_2CO}$. \cite{vantHoff2018} tentatively reported $\mathrm{D}_2\mathrm{CO}$ and $\mathrm{H}_2\mathrm{C^{18}O}$. We confirm the detection of $\mathrm{D_2CO}$ and tentatively detect $\mathrm{H_2C^{18}O}$.

\begin{table*}[!htbp]
\centering
\caption{HDCO and D$_2$CO transitions covered in this study from CDMS database}\label{HDCO_D2CO}
\begin{tabular}{lcc|lcc}
\hline \hline 
\multicolumn{3}{c|}{HDCO Transitions} & \multicolumn{3}{c}{D$_2$CO Detected Transitions} \\
\hline
Frequency (GHz) & $\mathrm{E_u}$ (K) & $\mathrm{A_{ul}\, (s^{-1})}$ & Frequency (GHz) & $\mathrm{E_u}$ (K) & $\mathrm{A_{ul}\, (s^{-1})}$ \\
\hline 
348.9650047 & 1017.78825 & $ 2.94 \times 10^{-5}$ & 349.63067 & 80.40747 & $ 1.08 \times 10^{-3}$ \\
350.2779245 & 2640.79933 & $ 2.99 \times 10^{-5}$ & 351.4916 & 145.249 & $ 7.04 \times 10^{-4}$ \\
351.2960031 & 1158.0205  & $ 1.06 \times 10^{-6}$ & 351.894  & 107.577 & $ 9.54 \times 10^{-4}$ \\
353.6786792 & 2181.69735 & $ 2.63 \times 10^{-6}$ & 352.2437 & 107.602 & $ 9.57 \times 10^{-4}$ \\
355.0753090 & 184.34584  & $ 6.08 \times 10^{-6}$ & 357.871  & 81.2117  & $ 1.19 \times 10^{-3}$ \\
361.799202  & 1047.58272 & $ 7.71 \times 10^{-7}$ &          &          &                       \\
363.1924917 & 2041.46266 & $ 1.54 \times 10^{-6}$ &          &          &                       \\
363.5958632 & 2041.46317 & $ 1.54 \times 10^{-6}$ &          &          &                       \\
364.7852433 & 717.47888  & $ 1.24 \times 10^{-6}$ &          &          &                       \\
364.8248437 & 136.63441  & $ 1.21 \times 10^{-6}$ &          &          &                       \\
\hline
\end{tabular}
\end{table*}

\begin{figure*}[!htbp]
\gridline{\fig{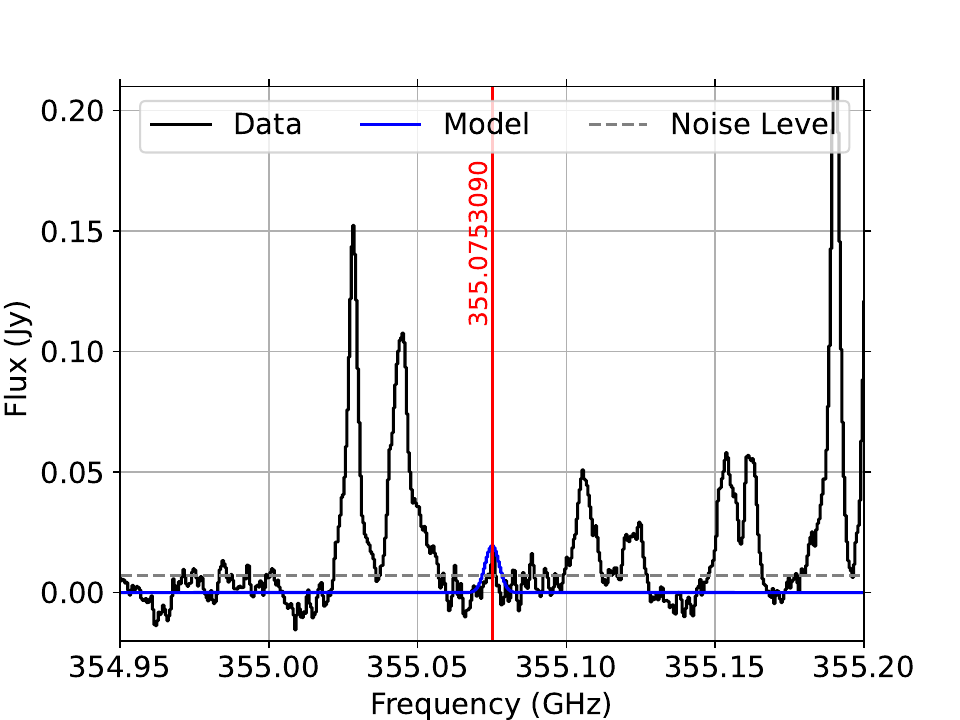}{0.45\textwidth}{}
          \fig{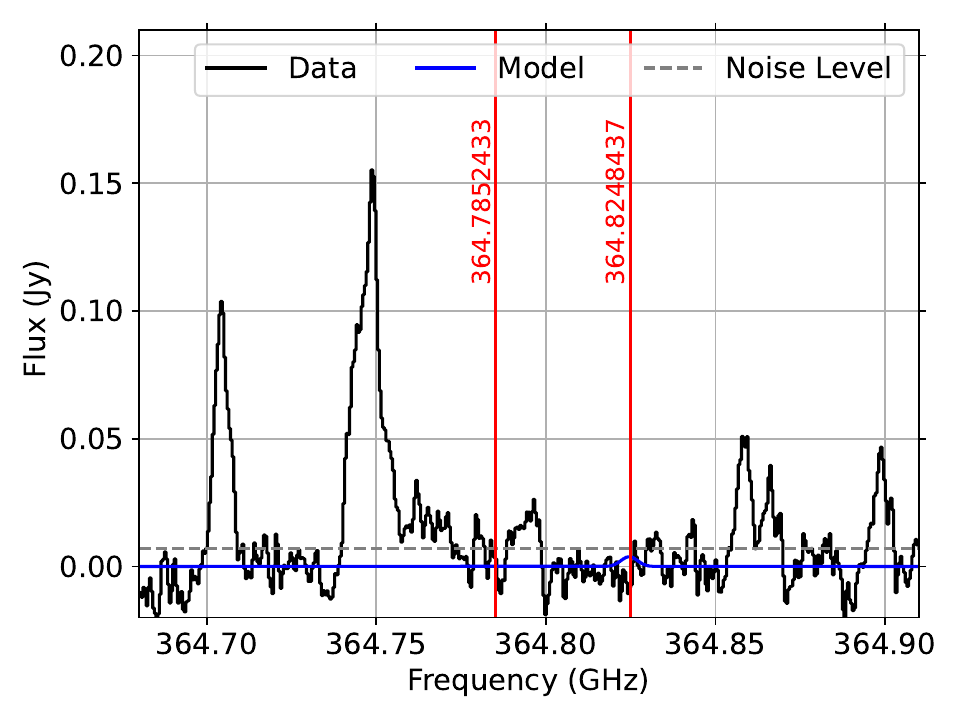}{0.4\textwidth}{}}
\caption{In blue is the simulated intensity at $\mathrm{T_{ex}} = 100$ K, and $\mathrm{N_{tot} = 3\times 10^{16}} \mathrm{cm^{-2}}$ of the three transitions of HDCO that are detectable under the physical conditions of this source. The red vertical line indicate the central frequency of each transition.
\label{HDCO}}
\end{figure*}

Despite identifying three isotopologues of $\mathrm{H_2CO}$, we did not detect the singly deuterated isotopologue $\mathrm{HDCO}$. \cite{lee2024ApJ} reported the detection of HDCO in this source at a frequency of 258.070936 GHz, with an $\mathrm{E_u}$ of 102.55758 K, and an $\mathrm{A_{ul}}$ of $2.11 \times 10^{-4}$ $\mathrm{s^{-1}}$. Our spectral scan covers 10 transitions of HDCO (see Table \ref{HDCO_D2CO}). Seven of these transitions have upper energy states ($\mathrm{E_u}$) $>$ 1000 K, and are thus unlikely to be detected. Figure \ref{HDCO} shows the simulated intensity of the three transitions of HDCO with $\mathrm{E_u}$ $<$ 1000 K.
Two of the lines are predicted to be below the noise level of our data. The the line at 355.0753 GHz is coincident with a weak feature at the level of the noise , insufficient to claim a detection according to the criteria outlined above.
In comparison, our spectral scan covers 15 transitions of $\mathrm{D_2CO}$; 7 of these transitions have $\mathrm{E_u}$ $>$ 1000 K, and all the transitions have $\mathrm{A_{ul}}$ values ranging between $10^{-6}$ and $10^{-3}; \mathrm{s^{-1}}$. We identify 5 transitions of $\mathrm{D_2CO}$ (Table \ref{HDCO_D2CO}), with $\mathrm{E_u}$ $>$ 150 K; hence, we consider the detection of $\mathrm{D_2CO}$ robust. The non-detection of HDCO is likely due to the strongest transitions being located outside our observed frequency range and not due to a low abundance of HDCO relative to $\mathrm{D_2CO}$.

\subsection{Glycolaldehyde \texorpdfstring{$(\mathrm{CH_2(OH)CHO})$}{CH2(OH)CHO}}

These observations were originally designed to detect the molecule $\mathrm{CH_2(OH)CHO}$ in this disk. Based on the abundance of $\mathrm{CH_2(OH)CHO}$ relative to $\mathrm{CH_3OH}$ in IRAS 16293-2422 (hereafter IRAS 16293) \citep{Manigand2020A&A}, we model the predicted flux in V883 Ori  assuming $T_{ex} = 103$ K and $N_{tot} = 4\times 10^{13}\, \mathrm{cm}^{-2}$. Assuming the same abundance ratio as in IRAS 16293, 18 transitions should be detected above $3\sigma$. However, the molecule was not detected, which indicates that $\mathrm{CH_2(OH)CHO}$ is not as abundant as in IRAS 16293. Even though our model predicted multiple transitions of this molecule, some are blended with other molecules while others do not match the data (see Appendix \ref{AppendixD}). This is consistent with the recent non-detection of $\mathrm{CH_2(OH)CHO}$ in Band 6 by \citet{Lee2025arXiv}. Despite the non-detection of $\mathrm{CH_2(OH)CHO}$, our spectral scan is rich in other COMs, including some that are being detected for the first time in this source. We also searched for additional COMs such as $\mathrm{\text{c-}C_2H_4NH}$, $\mathrm{CH_3NH_2}$, $\mathrm{CH_3COOH}$, $\mathrm{H_2NCH_2CN}$, $\mathrm{C_2H_3CN}$, $\mathrm{C_3H_5CN}$, $\mathrm{C_3H_7CN}$, $\mathrm{H_2CCNH}$, $\mathrm{HNCHCN}$, $\mathrm{C_2H_4S}$, $\mathrm{NH_2CH_2CH_2OH}$, $\mathrm{CH_2CN}$, $\mathrm{NH_2CH_2CN}$, $\mathrm{H_2CCNH}$, HOCN, and $\mathrm{\text{a\textquotesingle Gg-}(CH_2OH)_2}$, but none of these molecules were detected.

\section{Discussion} \label{discussion}

\begin{figure*}[!htbp]
\gridline{\fig{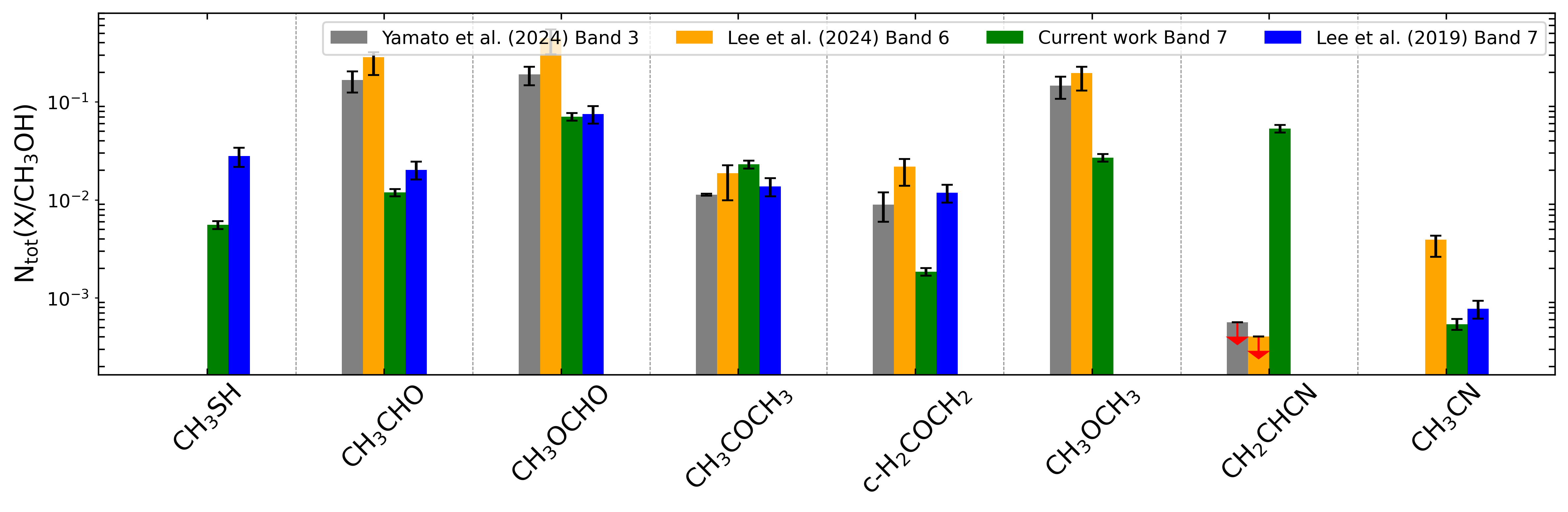}{1.0\textwidth\vspace{-10pt}}{}}
\caption{Comparison of various COM abundances relative to methanol in V883 Ori from this study with those in \cite{yamato2023chemistry} (Band 3 observations), \cite{Lee_2019} (Band 7 observations), and \cite{lee2024ApJ} (Band 6 observations). The red downward arrow indicates an upper limit column density.\label{v883comparison}}
\end{figure*}

V883 Ori's disk is one of the disks with the richest inventory of gas-phase COMs. This is due to the accretion outbursts that increase the temperature of the disk and sublimate the ices covering the COMs, allowing us to detect these molecules in the gas phase. Figure \ref{v883comparison} shows the comparison of the derived COMs abundances relative to $\mathrm{CH_3OH}$ in V883 Ori, observed with ALMA Band 3 \citep{yamato2023chemistry}, which covers the frequency range ($\sim$ 85 - 98 GHz), ALMA Band 7 \cite{Lee_2019}, which covers the frequency range ($\sim$ 335 - 350 GHz), ALMA Band 6 \cite{Lee2025arXiv}, which covers the frequency range ($\sim$ 220.7 - 274.9 GHz), and our current observations, which cover the frequency range ($\sim$ 348 - 366 GHz), including the first ALMA observations beyond 350 GHz in this source. In our observation, since the $\mathrm{CH_3OH}$ lines are optically thick, we used the $\mathrm{^{13}CH_3OH}$ column density multiplied by the $\mathrm{^{12}C}/\mathrm{^{13}C}$ ratio of 60 \citep{Langer93ApJ} to derive the ratios with respect to $\mathrm{CH_3OH}$, following the same method as \citet{Lee_2019}. 

For the Band 3 and Band 6 observations, the abundances relative to methanol were derived using the $\mathrm{CH_3OH}$ column density since the $\mathrm{CH_3OH}$ is optically thin in these observations. We report similar abundances compared to the previous Band 7 observations. For $\mathrm{CH_3SH}$ and $\mathrm{\text{c-}H_2COCH_2}$, only one transition was previously detected and the derived abundances are higher by an order of magnitude compared to those derived here.  We found ratios lower by an order of magnitude compared to Band 3 observations, except for the molecule $\mathrm{CH_3COCH_3}$, where we observed a similar ratio, and for the molecule $\mathrm{CH_2CHCN}$, where we report abundances higher by two orders of magnitude than Band 3. In both bands the line identification is considered tentative and there is insufficient data to robustly constrain the column density. We reported $\sim$ an order of magnitude lower abundances compared to Band 6 observations, except for the molecule $\mathrm{CH_3COCH_3}$, where similar abundances are reported. 

In general, the abundances of COMs in this work are more in agreement with the previous Band 7 observations, with only a small factor difference compared to Bands 3 and 6. These differences are likely due to a combination of factors. In particular, T$_{ex}$ was fixed for all molecules in the previous works, whereas in our model, we allowed T$_{ex}$ to vary. Additionally, different frequency ranges cover transitions with a different range of upper energies, and so may be probing different physical regions in the disk. The agreement in derived abundance relative to $\mathrm{CH_3OH}$ between our observations and the previous observations is particularly interesting, given the different T$_{ex}$ values. The abundance relative to $\mathrm{CH_3OH}$ for all detected molecules is reported in Table \ref{resultstable}, as well as the abundance relative to molecular hydrogen.

\begin{figure*}[!htbp]
\gridline{\fig{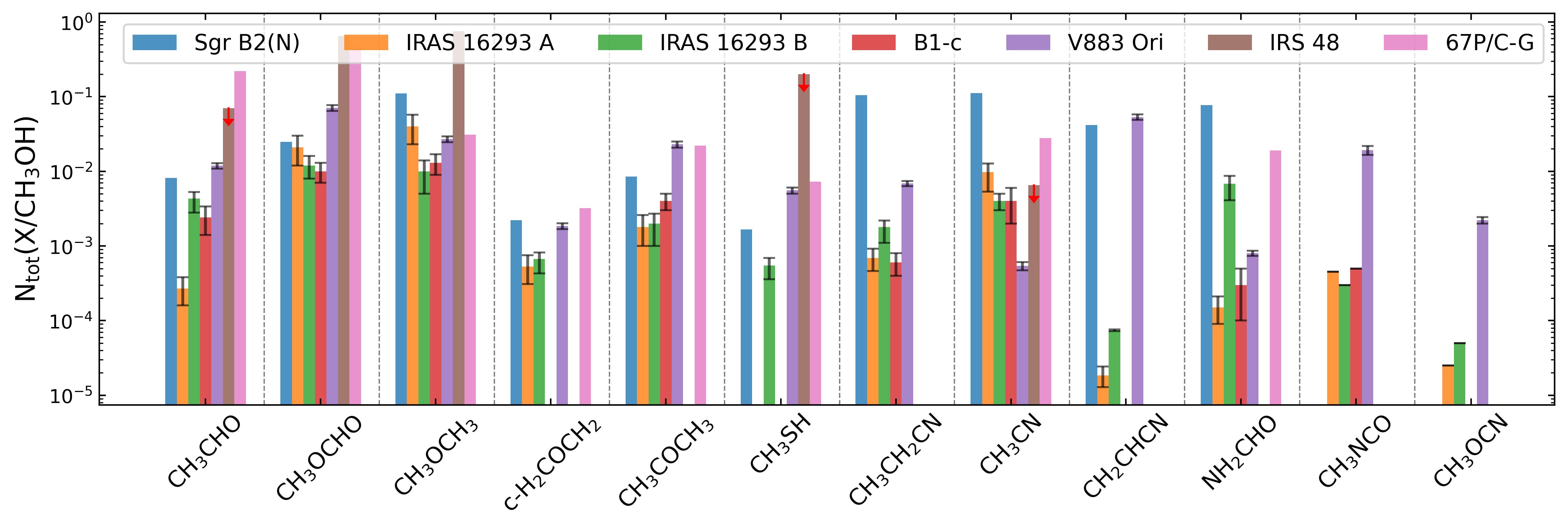}{1.0\textwidth\vspace{-10pt}}{}}
\caption{A comparison of the abundances of various COMs relative to $\mathrm{CH_3OH}$ across different sources, including V883 Ori, is presented in this work. The abundances relative to $\mathrm{CH_3OH}$ for the different sources were compiled from the literature as follows: for IRAS 16293 A and B, values are from \citet{Manigand2020A&A}, \citet{Calcutt2018A&A}, \citet{Manigand2019A&A}, and \citet{Nazari2024A&A}; for the protostar B1-c, from \citet{Gelder2020A&A} and \citet{Nazari2021A&A}; for the protoplanetary disk Oph-IRS 48, from \citet{Nashanty2022A&A};  for the comet 67P/C-G, from \citet{Hanni2023A&A} and \citet{Drozdovskaya2018MNRAS.476.4949D}, and for Sgr B2(N), from \cite{Belloche2013A&A}}. The red downward arrow in the source IRS 48 indicates the upper limit due to non-detection. \label{comparison}
\end{figure*}

Figure \ref{comparison} compares the abundances relative to $\mathrm{CH_3OH}$ in V883 Ori (current work) to different sources, including the Class 0 protostellar binary IRAS 16293 A and B (\citealt{Manigand2020A&A, Calcutt2018A&A, Manigand2019A&A, Nazari2024A&A}), the low mass protostar  B1-c (\citealt{Gelder2020A&A, Nazari2021A&A}), the protoplanetary disk around the Herbig Ae star Oph-IRS 48 \citep{Nashanty2022A&A}, the solar system comet 67P/C-G (\citealt{Hanni2023A&A, Drozdovskaya2018MNRAS.476.4949D}), and the high-mass star-forming region Sagittarius B2 (North) (hereafter Sgr B2(N)). We include Sgr B2 for completeness. Only a handful of sources have detections of these species. Comparing to an extreme environment gives insight into how common or rare the existence of these molecules is. For the source IRAS A, $\mathrm{CH_2CHCN}$, $\mathrm{CH_3CH_2CN}$, and $\mathrm{CH_3CN}$ column densities are from \citet{Calcutt2018A&A}, and the $\mathrm{CH_3OH}$ column density is from \citet{Manigand2020A&A}. For the source IRAS B, $\mathrm{CH_3CN}$ and $\mathrm{CH_2CHCN}$ column densities are from \citet{Calcutt2018A&A}, and the $\mathrm{CH_3OH}$ column density is from \citet{Jorgensen2018A&A}. Overall, the abundance ratios in V883 Ori tend to fall between those of the protostars and IRS48 and Comet 67P. We observed $\sim$1 to 3 orders of magnitude higher abundances of $\mathrm{CH_3CHO}$, $\mathrm{CH_3OCHO}$, $\mathrm{CH_3COCH_3}$, $\mathrm{CH_2CHCN}$, and $\mathrm{CH_3OCN}$ relative to $\mathrm{CH_3OH}$ compared to the Class 0 low-mass protostars IRAS 16293 A, IRAS 16293 B, and B1-c, except for $\mathrm{CH_3OCH_3}$, for which we report similar ratios. For $\mathrm{CH_3CN}$, we report an order of magnitude lower ratios compared to all sources. For $\mathrm{NH_2CHO}$ the ratio is similar to B1-c and IRAS 16293 A but an order of magnitude lower ratio compared to IRAS 16293 B. These differences could suggest that these molecules were thermally desorbed as more ices sublimated during the outburst of V883 Ori, given that these molecules have $T_{\text{ex}} \sim 70$--$100$ K. Additionally, we reported $\sim$ 1 to 3 orders of magnitude lower ratios compared to the protoplanetary disk of IRS 48, the solar system comet 67P/C-G, and Sgr B2 (N), except for $\mathrm{CH_3OCH_3}$, $\mathrm{\text{c-}H_2COCH_2}$, and $\mathrm{CH_3SH}$, where similar ratios were reported compared to 67P/C-G; for $\mathrm{CH_3CHO}$, $\mathrm{\text{c-}H_2COCH_2}$, and $\mathrm{CH_2CHCN}$, where we reported similar ratios compared to Sgr B2 (N); and for $\mathrm{CH_3OCHO}$, $\mathrm{CH_3COCH_3}$, and $\mathrm{CH_3SH}$, where $\sim$ 2 - 3 times higher ratios are reported compared to Sgr B2 (N). These results suggest that chemical complexity could increase during the protoplanetary disk phase. Whether this can be expected in all disks, or only those such as V883 Ori which experience significant accretion bursts, remains unclear. The combination of increased desorption and additional gas phase reactions could produce changes in the chemical composition of disks with and without a history of FU Ori-like outbursts.


\begin{figure*}[!htbp]
\gridline{\fig{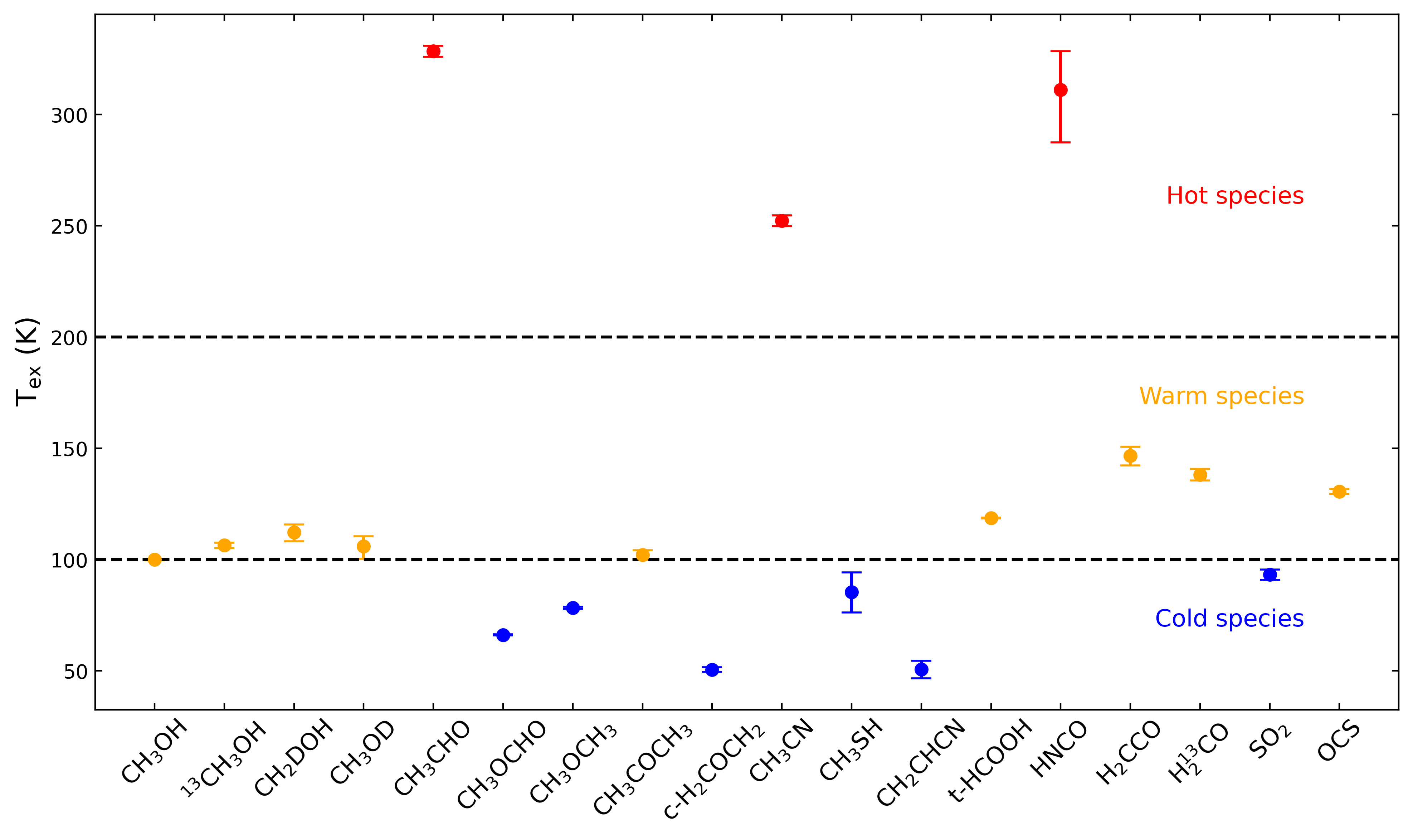}{1.0\textwidth\vspace{-10pt}}{}}
\caption{The three categories} of T$_{ex}$ for different molecular species. The first category includes molecules with low T$_{ex}$, shown in blue (T$_{ex}$ $\sim$ 50 - 100 K). The second category represents warm species (T$_{ex}$ $\sim$ 100 - 200 K), shown in orange, and the third includes hot molecules (T$_{ex}$ $\sim$ 200 - 400 K), shown in red. We only plotted molecules for which T$_{ex}$ is well constrained by the model. \label{temperature}
\end{figure*}

\subsection{Excitation temperature variation}
Previous observations of V883 Ori (\citealt{Lee_2019, yamato2023chemistry, Lee2025arXiv}) assumed the excitation temperature to be common for all molecules, while in our model, excitation temperature is a free parameter. We reported different excitation temperature that can be classified into three general categories: those with excitation temperatures near the sublimation temperature of H$_\mathrm{2}$O ($\sim 115$ K) and those with much higher or lower excitation temperatures. Most of the observed species have excitation temperatures similar to the sublimation temperature of H$_\mathrm{2}$O, suggesting these species sublimated from the grain surface alongside H$_\mathrm{2}$O. This has been previously observed for, e.g., CH$_3$OH, which has a similar grain surface binding energy to H$_\mathrm{2}$O \citep{vantHoff2018,Tobin2023}. $\mathrm{CH_3CHO}$, $\mathrm{CH_3CN}$, and $\mathrm{HNCO}$  have much higher temperatures. These molecules do not have significantly higher binding energies than, e.g.,  $\mathrm{CH_3OH}$ \citep{Wakelam2017}. The higher excitation temperatures are therefore not indicative of a higher sublimation temperature. It is likely this emission originates from a range of radii, and thus, temperatures, while being dominated by the hot inner disk. For an out-bursting source like V883 temperatures can reach several hundred Kelvin, even out to 20 au \citep{Alarcon2024}. There are also several COMs with  temperatures closer to the $\mathrm{CO_2}$ sublimation temperature ( $\sim$ 50~K): $\mathrm{\text{c-}H_2COCH_2}$, $\mathrm{CH_2CHCN}$, $\mathrm{CH_3OCHO}$ among others. 

Figure \ref{temperature} shows the three classifications of different molecules whose T$_{ex}$ values are well constrained by the model. We report similar T$_{ex}$ values to those observed in high- and low-mass protostars for some molecules (\citealt{Nazari2022A&A, Jorgensen2018A&A, Nazari2024A&A}). For $\mathrm{CH_3CHO}$, we reported a temperature $\sim$ 200 K higher compared to (\citealt{Jorgensen2018A&A, Nazari2024A&A}), and for $\mathrm{CH_3OCHO}$, we reported a temperature $\sim$ 200 K lower compared to \citet{Jorgensen2018A&A}. In the low-mass protostar IRAS 16293 B (\citealt{Jorgensen2018A&A, Nazari2024A&A}), a similar variation in T$_{ex}$ is reported. However, \citet{Jorgensen2018A&A} suggested that this variation could be related to the binding energy of the molecule. This is not the case in our observations, as it is challenging to observe a clear pattern between the binding energy and T$_{ex}$ of some molecules. Both cold and warm species have binding energies ranging from $\sim$ 2000 to 7000 K. Molecules with low binding energy are expected to desorb at lower T$_{ex}$ ($\sim$ 50 K), and molecules with high binding energy are expected to desorb at higher T$_{ex}$ ($\sim$ 100 K) if the molecule is bound in ices. If the molecule is mixed with $\mathrm{CH_3OH}$ or $\mathrm{H_2O}$, its binding energy will increase, leading it to desorb at the T$_{ex}$ of $\mathrm{CH_3OH}$ or $\mathrm{H_2O}$ \citep{Jorgensen2018A&A}. However, some molecules, such as $\mathrm{\text{c-}H_2COCH_2}$ and $\mathrm{CH_2CHCN}$, have high binding energies ($\sim$ 4000 K), but their T$_{ex}$ is $\sim$ 50 K; suggesting they were not produced via direct sublimation from the ice. The hot species have binding energies of $\sim$ 2700 to 4700 K, these molecules are likely to originate from the innermost region. Furthermore, these species could also be formed through the gas-phase reactions (\cite{Charnley1992, Garrod2008ApJ}).

\subsection{Nitrogen-bearing COMs}
Previous observations of V883 Ori showed that there is a deficiency in detected nitrogen-bearing COMs. The only N-bearing COM detected to date in this disk is $\mathrm{CH_3CN}$ (e.g., \citealt{Lee_2019, Lee2025arXiv}). We also detect $\mathrm{CH_3CN}$ and newly detect $\mathrm{CH_3CH_2CN}$, with multiple tentative detections of N-bearing COMs such as $\mathrm{NH_2CHO}$, $\mathrm{CH_3OCN}$, $\mathrm{CH_3NCO}$, and $\mathrm{CH_2CHCN}$. \cite{Garrod2008ApJ} suggested that N-bearing species could form through the combination of radicals formed mainly by cosmic-ray-induced photodissociation. \cite{Halfen2015ApJ} proposed that the formation route of $\mathrm{CH_3NCO}$ could be via HNCO or HOCN in the gas phase through reactions with $\mathrm{CH_3}$:
\begin{equation}
    \mathrm{HNCO} + \mathrm{CH_3} \longrightarrow \mathrm{CH_3NCO} + \text{H}
\end{equation}
\begin{equation}
    \mathrm{HOCN} + \mathrm{CH_3} \longrightarrow \mathrm{CH_3NCO} + \text{H}
\end{equation}
\noindent There are other formation routes for $\mathrm{CH_3NCO}$, but in all cases, it is required that the reactants are sufficiently available. Additionally, \cite{Garrod2008ApJ} suggested that some COMs can be formed via functional group substitution. For example, the $\mathrm{NH_2}$ radical substitutes the aldehyde group (-CHO) in the $\mathrm{CH_3CHO}$, which leads to the formation of $\mathrm{NH_2CH_3}$ (methylamine), with a formyl radical (HCO) left over from the reaction.
\begin{equation}
    \mathrm{CH_3CHO} + \mathrm{NH_2} \longrightarrow \mathrm{NH_2CH_3} + \text{HCO}
\end{equation}
\noindent On the other hand, instead of substituting the aldehyde group, $\mathrm{NH_2}$ substitutes the $\mathrm{CH_3}$, which lead to the formation of $\mathrm{NH_2CHO}$ (formamide) and $\mathrm{CH_3}$ (methyl radical).
\begin{equation}
    \mathrm{CH_3CHO} + \mathrm{NH_2} \longrightarrow \mathrm{NH_2CHO} + \mathrm{CH_3}
\end{equation}
\noindent Our results showed an abundance of HNCO/$\mathrm{CH_3NCO}$ = 0.18, which is an order of magnitude lower compared to the low-mass protostars IRAS 16293 A and B ($\sim$ 4$-$12, \citealt{Ligterink2017MNRAS}). We also reported an abundance of $\mathrm{CH_3NCO}$/$\mathrm{CH_3OCN}$ $>$ 8, which is similar to low-mass protostars IRAS 16293 A and B ($>$ 10, \citealt{Ligterink2017MNRAS}). $\mathrm{CH_3NCO}$ and $\mathrm{NH_2CHO}$ are considered prebiotic precursors, as they could play a role in the synthesis of amino acid chains \citep{Pascal}. The formation of vinyl cyanide ($\mathrm{CH_2CHCN}$) and ethyl cyanide ($\mathrm{CH_3CH_2CN}$) are closely linked. \citep{Garrod2017A&A} suggested that the formation route of $\mathrm{CH_2CHCN}$ takes place during the collapse stage on the dust-grain surfaces through the hydrogenation of cyanoacetylene ($\mathrm{HC_3N}$). However, $\mathrm{CH_2CHCN}$ is often further hydrogenated on the dust grains into $\mathrm{CH_3CH_2CN}$ as shown in the following reaction:
\begin{equation}
    \mathrm{H} + \mathrm{HC_3N} \longrightarrow \mathrm{C_2H_2CN} 
\end{equation}
\begin{equation}
    \mathrm{H} + \mathrm{C_2H_2CN} \longrightarrow \mathrm{C_2H_3CN} 
\end{equation}
\begin{equation}
    \mathrm{H} + \mathrm{C_2H_3CN} \longrightarrow \mathrm{CH_2CH_2CN/CH_3CHCN} 
\end{equation}
\begin{equation}
    \mathrm{H} + \mathrm{CH_2CH_2CN/CH_3CHCN} \longrightarrow \mathrm{C_2H_5CN} 
\end{equation}
\noindent Furthermore, \citep{Garrod2017A&A} suggested that the hydrogenation process results in low $\mathrm{CH_2CHCN}$ abundances during the cold phase because most of it converted into $\mathrm{CH_3CH_2CN}$. During the warm-up phase, $\mathrm{CH_3CH_2CN}$ is destroyed through protonation and dissociative recombination, producing $\mathrm{CH_2CHCN}$, which increases the abundance of $\mathrm{CH_2CHCN}$ in the gas phase. We reported a $\mathrm{CH_2CHCN}$/$\mathrm{CH_3CH_2CN}$ ratio of $\sim 21$, which strongly agrees with the prediction of \citet{Garrod2017A&A} that $\mathrm{CH_2CHCN}$ abundance will increase during the warm-up phase.

A similar difference between O-bearing COMs and N-bearing COMs has also been observed towards the Orion hot core and Compact Ridge. A higher abundance of N-bearing than O-bearing COMs has been observed towards the hot core, and the opposite is observed towards the Compact Ridge \citep{Blake1987ApJ}. Several chemical models have been used to explain these differences (\citealt{Charnley1992, Caselli1993ApJ}). \cite{Caselli1993ApJ} attributes these differences to the interaction of radiation, energetic stellar winds, and outflows with the surrounding environment, which heat the two regions, leading to different chemistry. Another possible scenario that could explain this discrepancy is the presence of shocks, which could enhance the O-bearing COMs in the Compact Ridge (see, e.g., \cite{Blake1987ApJ}). Furthermore, this discrepancy between the N- and O-bearing COMs has also been observed towards the G35.20-0.74N and G35.03+0.35 hot cores \citep{Allen2017A&A}, and are attribute it to several factors, including temperature differences. They find that N-bearing species tend to be more abundant in hotter regions. These scenarios could explain the deficiency seen in V883 Ori, especially the presence of the shock, as this source is undergoing an accretion outburst that enhances O-bearing COMs while, at the same time, potentially destroying the N-bearing COMs \citep{Zwicky2024MNRAS}. 

\begin{figure*}[!htbp]
\gridline{\fig{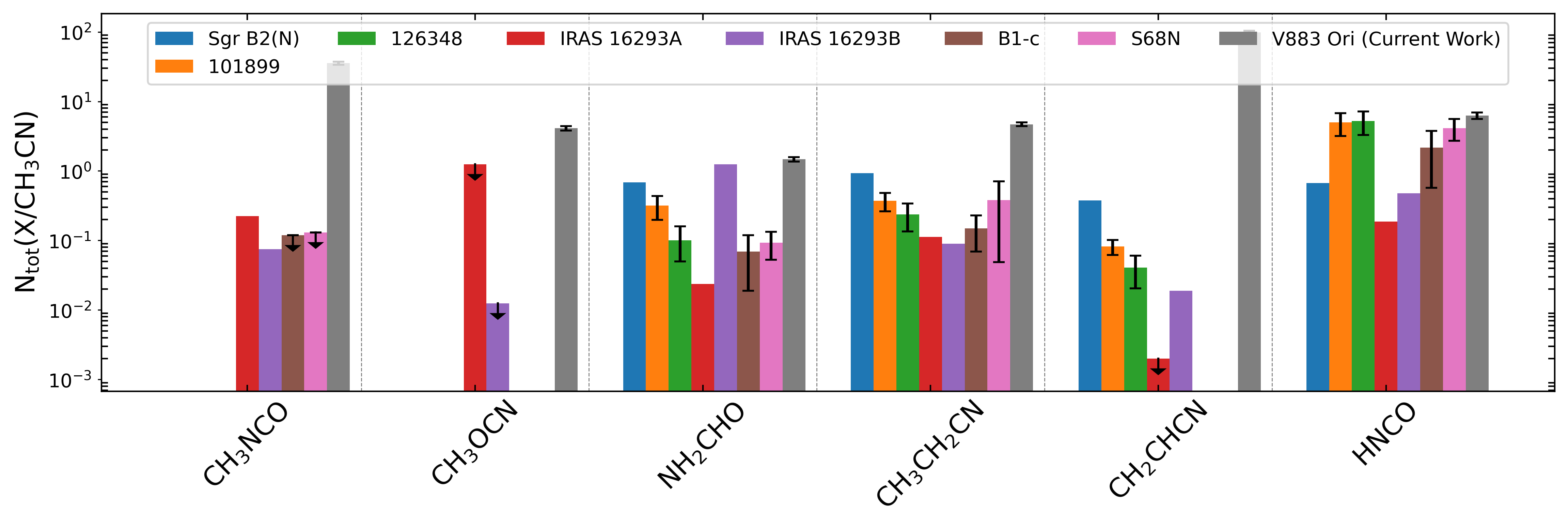}{1.0\textwidth\vspace{-10pt}}{}}
\caption{Comparison of N-bearing species relative to $\mathrm{CH_3CN}$ in low- and high-mass protostars. The abundances are compiled from the literature as follows: Sag B2 (N) from \cite{Belloche2013A&A}, 101899 and 126348 are from \cite{Nazari2022A&A}, B1-c and S68N are from \cite{Nazari2021A&A}, and IRAS 16293 A and B are from \cite{Calcutt2018A&A, Manigand2020A&A, Ligterink2017MNRAS}, and \cite{Zeng2019MNRAS}. The black down arrow indicates the upper limit due to tentative detection. For HNCO, $\mathrm{NH_2CHO}$, $\mathrm{CH_3OCN}$ and $\mathrm{CH_3NCO}$ in IRAS 16293 A, the column density of $\mathrm{CH_3CN}$ is adopted from \cite{Calcutt2018A&A}.
\label{NCH3CN}}
\end{figure*}

\begin{figure*}[!htbp]
\gridline{\fig{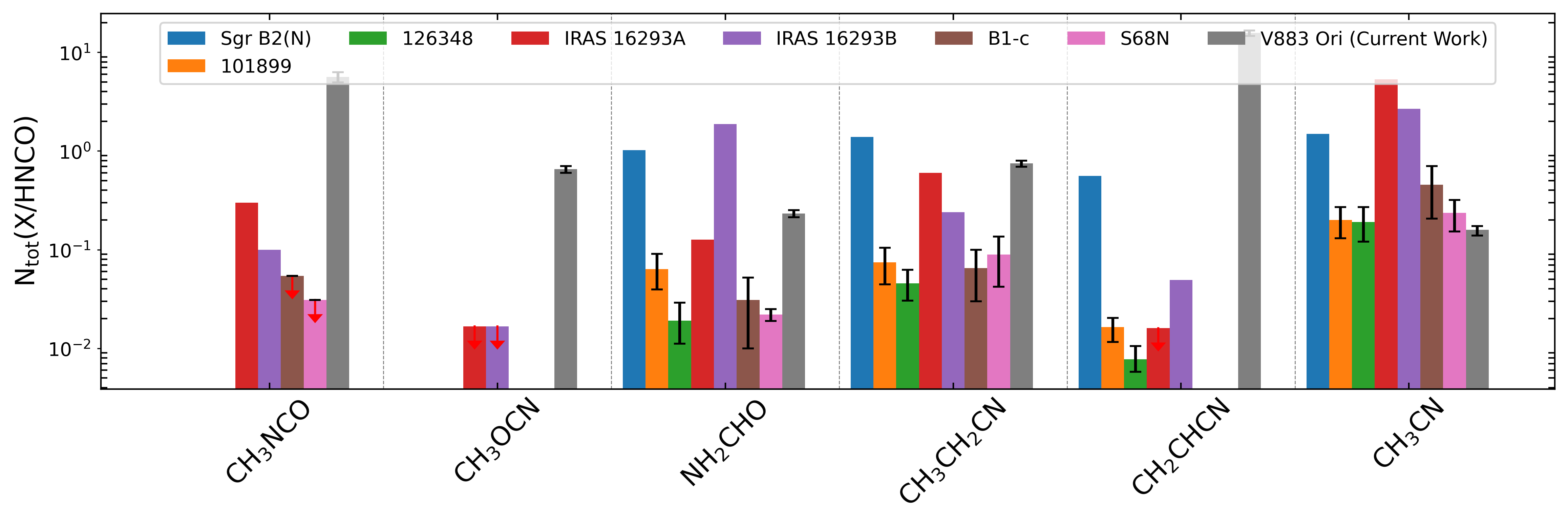}{1.0\textwidth\vspace{-10pt}}{}}
\caption{Comparison of N-bearing species relative to HNCO in low- and high-mass protostars. The abundances are compiled from the literature as follows: Sag B2 (N) from \cite{Belloche2013A&A}, 101899 and 126348 are from \cite{Nazari2022A&A}, B1-c and S68N are from \cite{Nazari2021A&A}, and IRAS 16293 A and B are from \cite{Calcutt2018A&A, Manigand2020A&A, Ligterink2017MNRAS}, and \cite{Zeng2019MNRAS}. The red down arrow indicates the upper limit due to tentative detection. For $\mathrm{CH_3CH_2CN}$, $\mathrm{CH_2CHCN}$ and $\mathrm{NH_2CHO}$ in IRAS 16293 A and B, the column density of $\mathrm{HNCO}$ is adopted from \cite{Manigand2020A&A}. For $\mathrm{HOCH_2CN}$ in IRAS 16293 B, the column density of $\mathrm{HNCO}$ is adopted from \cite{Ligterink2017MNRAS}}.
\label{NHNCO}
\end{figure*}

Figure \ref{NCH3CN} and \ref{NHNCO} show the abundances of nitrogen-bearing species relative to $\mathrm{CH_3CN}$ and HNCO for the low-mass protostars B1-c, S68N, IRAS 16293 A, and IRAS 16293 B (\citealt{Nazari2021A&A, Calcutt2018A&A, Manigand2020A&A, Ligterink2017MNRAS, Zeng2019MNRAS}), the high-mass protostars 101899 and 126348 \citep{Nazari2022A&A}, and Sgr B2(N) \citep{Garrod2017A&A}. We reported $\sim$ 1 to 5 orders of magnitude higher ratio in V883 Ori relative to $\mathrm{CH_3CN}$ compared to all sources for most of the molecules. For $\mathrm{HNCO}$, we reported a similar ratio to 101899, 126348, B1-c, and S68N, and for the molecule $\mathrm{NH_2CHO}$, a similar ratio to IRAS 16293 B. Additionally, we reported a $\sim$ 1 to 3 orders of magnitude higher ratio relative to $\mathrm{HNCO}$ compared to all sources for most of the molecules. For the molecule $\mathrm{CH_3CN}$, we reported a similar ratio compared to 101899, 126348, and S68N, and a lower ratio compared to B1-c, IRAS 16293 A, and IRAS 16293 B. We also reported a lower ratio for the molecule $\mathrm{NH_2CHO}$ relative to Sgr B2(N) and IRAS 16293 B. For 67P the $\mathrm{CH_3NCO/CH_3CN}$ and $\mathrm{CH_3NCO/HNCO}$ ratios are $\sim$ 4 \citep{Goesmann2015Sci}. For V883 Ori, we find an 8-times higher ratio of the molecule $\mathrm{CH_3NCO}$ relative to $\mathrm{CH_3CN}$ and a consistent ratio relative to $\mathrm{HNCO}$. These results suggest that N-bearing COMs are enhanced during the transition from the protostellar envelope to the protoplanetary disks, though additional processing occurs before the final comet composition is set. Additionally, the reported column densities of $\mathrm{CH_3CN}$ and $\mathrm{HNCO}$ are $\sim$ 1 $-$ 2 orders of magnitude lower compared to other N-bearing COMs, which leads to higher relative abundances of these molecules, suggesting chemistry has converted nitrogen into more complex species between the protostellar and protoplanetary disk stage. However, most of the N-bearing COMs reported in Figures \ref{NCH3CN} and \ref{NHNCO} are tentative detections and, therefore, their column density are uncertain. Further observations of N-bearing COMs in the V883 Ori disk, as well as in other protoplanetary disks, are required to support these results. 

\subsection{Limitations of Derived Abundances}

In this section, we explore different sources of uncertainty that could affect the measurement of the column density of each molecule and, consequently, the derived relative molecular abundances.
Firstly, dust opacity can significantly attenuate the line emission of molecules residing in the innermost region, as has already been demonstrated in previous observations of this source (\citealt{vantHoff2018, Lee_2019, yamato2023chemistry}). In addition, \cite{DeSimone2020ApJ} find that dust absorption at millimeter wavelengths can significantly reduce molecular line intensities for optically thin lines, leading to an underestimation of molecular abundances and a deficiency in detected COMs.
In (sub)millimeter observations, bright dust emission can often obscure molecular emission, especially for molecules residing in the disk midplane. In this source, the dust optical depth at the disk midplane ($0.1\arcsec \simeq$ 42 au) could reach $\gtrsim$ 2 \citep{Cieza2016Natur.535..258C}, which could significantly underestimate the molecular abundances originating from this region.


Secondly, the $\mathrm{^{12}C/^{13}C}$ ratio in star-forming regions has been found to be significantly variable compared to the ISM (\citealt{Jorgensen2018A&A, yamato2023chemistry, Hsieh2025}). In V883 Ori, \cite{yamato2023chemistry} reported a ratio of $\mathrm{^{12}C/^{13}C} = 20 - 30$ for multiple COMs, which is significantly lower than the ISM value ($\sim$ 62; \cite{Langer93ApJ}, and $\sim$ 68; \cite{Milam2005ApJ}). This low ratio can be explained by isotopic fractionation processes \citep{Colzi2020A&A}. 
Furthermore, \cite{Yoshida2022ApJ} reported a lower $\mathrm{^{12}CO/^{13}CO}$ ratio in the innermost region and a higher ratio in the outer disk of TW Hya. They suggested that in low-temperature regions and high C/O ratios ($>$ 1), where isotope exchange reactions take place (see, e.g., \cite{Langer1984ApJ}), a lower $\mathrm{^{12}CO/^{13}CO}$ ratio could occur, especially in the innermost region. Moreover, the $\mathrm{^{12}C/^{13}C}$ ratio is also affected by whether the molecule is formed from carbon atoms or from CO molecules. The former will have a higher ratio compared to the canonical $\mathrm{^{12}C/^{13}C}$, while the latter will have a lower ratio, as demonstrated by \cite{Furuya2011ApJ...731...38F}.
Hence, these variations in the $\mathrm{^{12}C/^{13}C}$ ratio could lead to significant uncertainty when using the canonical ISM value of $\mathrm{^{12}C/^{13}C}$ to obtain a $\mathrm{CH_3OH}$ column density from $\mathrm{^{13}CH_3OH}$ in the comparison of molecular abundances across different evolutionary stages.

Thirdly, fixing the emitting area of each molecule to the same value could impact the derived molecular abundances, as has been shown by a recent study \citep{Hsieh2024A&A}. To check the validy of this assumption, we calculate the FWHM for several molecular transitions before applying kinematical corrections to the spectra (see, e.g., Appendix~\ref{AppendixE}). The observed FWHM values range from 2.69 km s$^{-1}$ (HNC) to 5.42 km s$^{-1}$ (CH$_3$CHO). A larger FWHM corresponds to higher velocity emission, which indicates an origin from smaller disk radii. The COMs (CH$_3$OH, CH$_3$CHO, CH$_3$OCH$_3$) show the largest FWHM values, implying that their emission mainly comes from the hotter inner disk. This is expected, as these species have higher binding energies and only sublimate into the gas phase in regions where the disk temperature is sufficiently high. Thus, the assumption of a fixed source size could introduce uncertainties in the relative molecular abundances, especially when different molecules trace different regions of the disk.

\section{Conclusions} \label{sec:conclusion}

We present in this paper ALMA Band 7 observations of V883 Ori. We detected 14 COMs, including isotopologues: $\mathrm{CH_3OH}$, $\mathrm{^{13}CH_3OH}$, $\mathrm{CH_2DOH}$, $\mathrm{CH_3OD}$, $\mathrm{CH_3 ^{18}OH}$, $\mathrm{CH_3OCH_3}$, $\mathrm{\text{c-}H_2COCH}_2$,  $\mathrm{CH_3OCHO}$, $\mathrm{CH_3SH}$, $\mathrm{CH_3CHO}$, $\mathrm{^{13}CH_3CHO}$, $\mathrm{CH_3 ^{13}CHO}$, $\mathrm{CH_3COCH_3}$ and $\mathrm{CH_3CN}$. We also identified 12 other molecules, including isotopologues:  $\mathrm{SO_2}$, $\mathrm{HCN}$, $\mathrm{HNC}$, $\mathrm{H_2CO}$, $\mathrm{H_2 ^{13}CO}$, $\mathrm{H_2C^{17}O}$, $\mathrm{D_2CO}$, $\mathrm{HNCO}$, $\mathrm{HCO^+}$,  $\mathrm{H_2CCO}$, $\mathrm{OCS}$, and $\mathrm{\text{t-}HCOOH}$. We detect three species for the first time in this source: $\mathrm{CH_3OD}$, $\mathrm{H_2C ^{17}O}$, and $\mathrm{H_2 ^{13}CO}$. Further, we confirm the presence of $\mathrm{CH_3SH}$, and 
$\mathrm{D_2CO}$ which are tentatively reported in the previous studies. We also tentatively reported: $\mathrm{NH_2CHO}$, $\mathrm{H_2C^{18}O}$, $\mathrm{CH_3NCO}$, $\mathrm{CH_3OCN}$, CH$_3$CH$_2$CN, and $\mathrm{CH_2CHCN}$. We derived the abundance relative to $\mathrm{CH_3OH}$ and compared the abundance to previous work on the same source using ALMA Band 3, 6, and 7 observations (see Figure \ref{v883comparison}). We report similar abundances compared to Band 7 observations within a factor of a few. We reported an order of magnitude lower abundances compared to Band 3 and 6 observations for some molecules, while others showed similar and low abundances. Additionally, we compared the abundance relative to $\mathrm{CH_3OH}$ for COMs to Class 0 protostars of IRAS 16293 A, IRAS 16293 B, and B1-c, the protoplanetary disk of IRS 48, Sgr B2 (N), and the solar system comet 67P/C-G (see Figure \ref{comparison}). We report $\sim$ 1 to 3 order of magnitude higher abundances of various COMs relative to $\mathrm{CH_3OH}$ compared to the Class 0 protostars of IRAS 16293 A and B, and B1-c. This suggests that molecular abundances are enhanced during the evolution from the protostellar envelope to protoplanetary disks. Additional chemical modeling studies are needed to determine if the current active state of V883 Ori has significantly altered the dominant chemical pathways in the disk or primarily increased the desorption of pre-existing ices. Additionally, we reported $\sim$ 1 to 3 order of magnitude lower ratios compared to the protoplanetary disk of IRS 48 and the solar system comet 67P/C-G. Furthermore, We found that the N-bearing COMs are still deficient in this disk; we detected only two N-bearing COMs and reported a tentative detection of multiple other species. Additional observations are required, especially at longer wavelengths where the dust emission is lower compared to Band 7, to confirm the detection of these molecules and to further identify new N-bearing COMs.

\begin{acknowledgments}
We thank the anonymous referee for their constructive comments and suggestions, which have helped improve the quality of this paper. This paper makes use of the following ALMA data: ADS/JAO.ALMA\#2021.1.00452.S. ALMA is a partnership of ESO (representing its member states), NSF (USA) and NINS (Japan), together with NRC (Canada) and NSC and ASIAA (Taiwan), in cooperation with the Republic
of Chile. The Joint ALMA Observatory is operated by ESO, AUI/NRAO and NAOJ. 

K.S. and T.S. would like to acknowledge European Research Council under the Horizon 2020 Framework Program via the ERC Advanced Grant Origins 83 24 28.

\end{acknowledgments}

%

\vspace{5mm}
\facilities{ALMA}


\software{Astropy \citep{2013A&A...558A..33A,2018AJ....156..123A, astropy:2022},  Gofish \citep{GoFish}, Emcee \citep{emcee2013}, CASA \citep{The_CASA_Team_2022}, Numpy \citep{numpyharris2020array}, Matplotlib \citep{matplotlibHunter:2007}, corner.py \citep{corner}, Pandas \citep{pandasmckinney-proc-scipy-2010}
}

\section*{ORCID iDs}

\noindent Abubakar M. A. Fadul \orcidlink{0009-0003-6626-8122}: \url{https://orcid.org/0009-0003-6626-8122}

\noindent Kamber R. Schwarz \orcidlink{0000-0002-6429-9457}: \url{https://orcid.org/0000-0002-6429-9457}

\noindent Jane Huang \orcidlink{0000-0001-6947-6072}: \url{https://orcid.org/0000-0001-6947-6072}

\noindent Jennifer B. Bergner \orcidlink{0000-0002-8716-0482}: \url{https://orcid.org/0000-0002-8716-0482}

\noindent Jenny K. Calahan \orcidlink{0000-0002-0150-0125}: \url{https://orcid.org/0000-0002-0150-0125}

\noindent Merel L. R. van ’t Hoff \orcidlink{0000-0002-2555-9869}: \url{https://orcid.org/0000-0002-2555-9869}

\noindent Tushar Suhasaria \orcidlink{0000-0002-4755-4719}: \url{https://orcid.org/0000-0002-4755-4719}

\bibliography{sample631}{}
\bibliographystyle{aasjournal}
\appendix



\section{Appendix A}\label{AppendixA}

Figure~\ref{spec}, \ref{spec1}, \ref{spec2}, \ref{spec3}, \ref{spec4}, \ref{spec5}, \ref{spec6}, and \ref{spec7} represent a zoom-in of the modeled spectra overlaid on the observed spectra. In black is the observed data, and different colors represent the best fit from the model of different molecular species. The dashed horizontal line in gray represents the noise level in the data (1$\sigma$).

\begin{figure}[!htbp]
    \centering
    \includegraphics[width=0.85\textwidth]{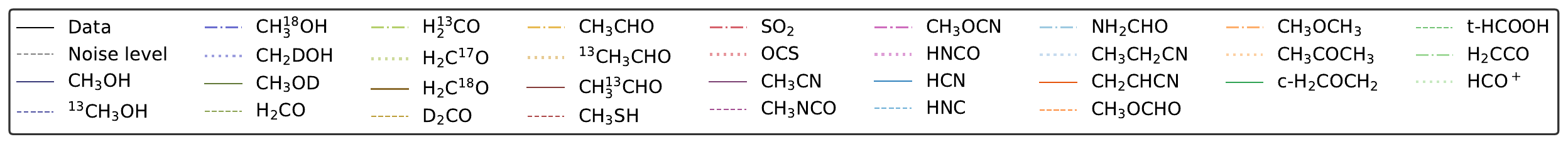} \\
    \includegraphics[width=0.85\textwidth]{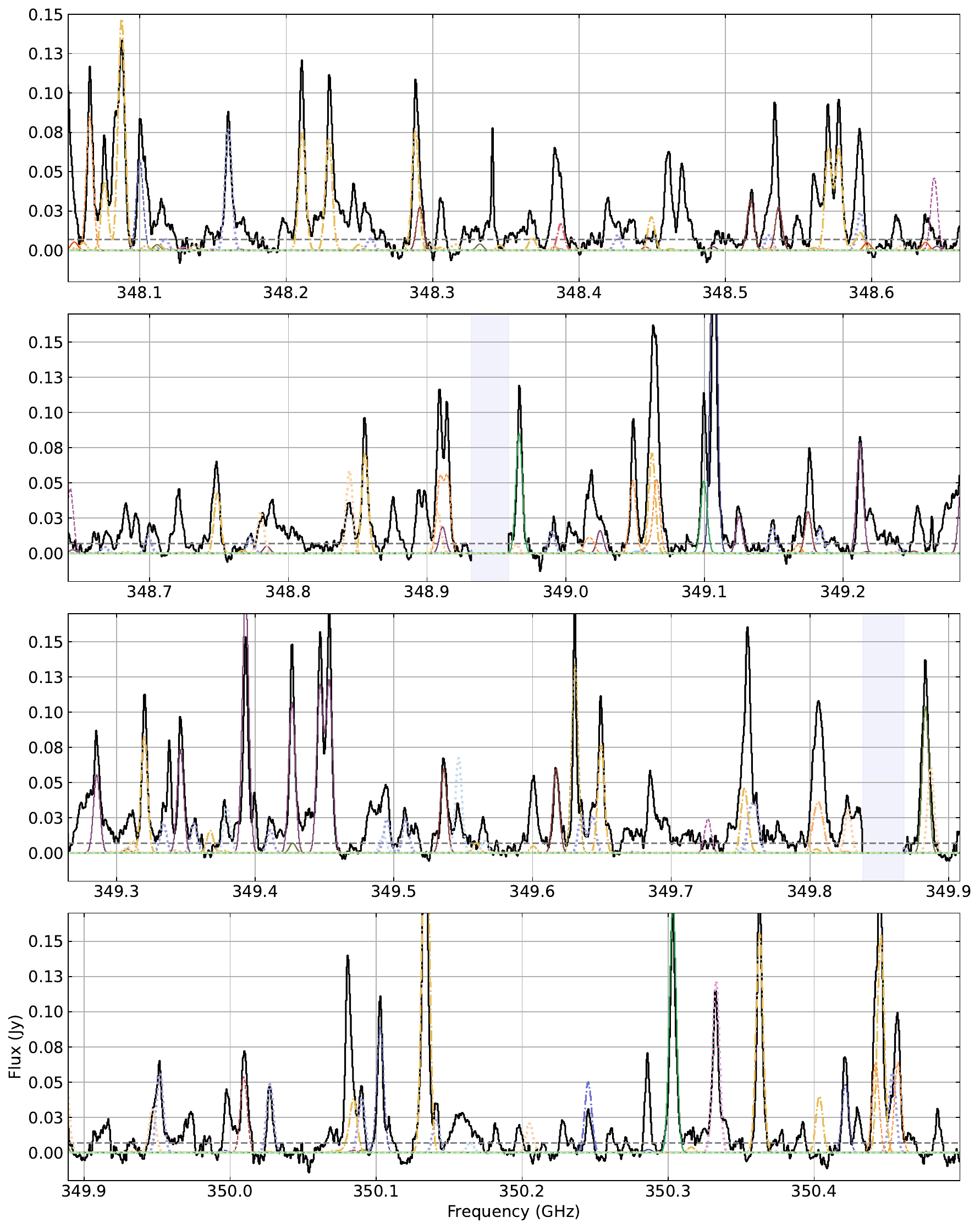}
    \caption{Detected COMs towards the disk of V883 Ori. The horizontal dotted line in gray represent the noise level ($1\sigma$) in the data, the black line represent the data, and different colors represent different species identified in this study, the dashed lines to facilitate distinguishing similar colors. The gaps between the spectral windows are excluded from the fit and shaded in lavender.
    \label{spec}}
\end{figure}
\clearpage

\begin{figure}[!htbp]
    \centering
    \includegraphics[width=0.9\textwidth]{figures/molecules_legend.pdf} \\
    \includegraphics[width=0.9\textwidth]{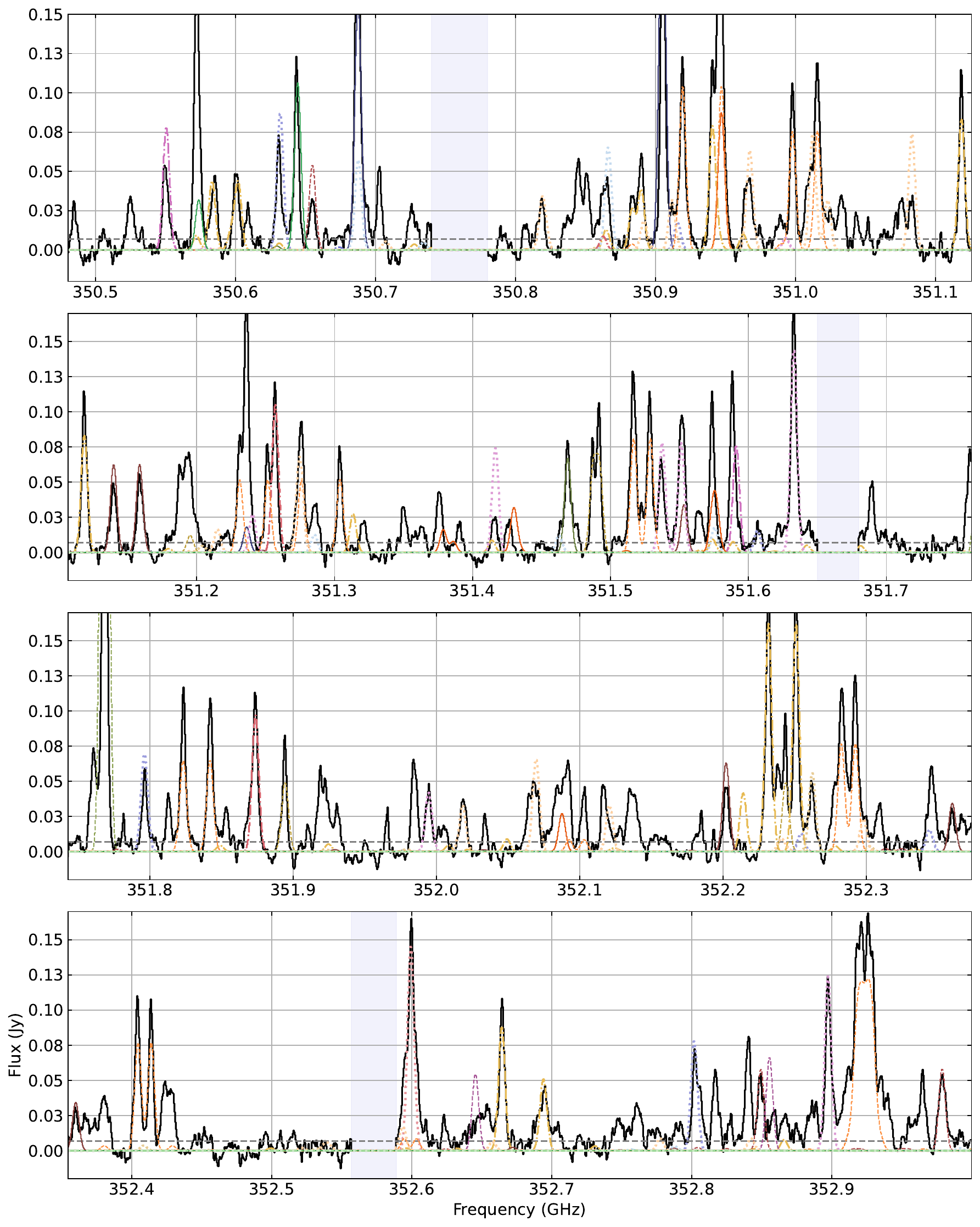}
    \caption{Continuation of figure \ref{spec}.\label{spec1}}
\end{figure}

\begin{figure}[!htbp]
    \centering
    \includegraphics[width=0.9\textwidth]{figures/molecules_legend.pdf} \\
    \includegraphics[width=0.9\textwidth]{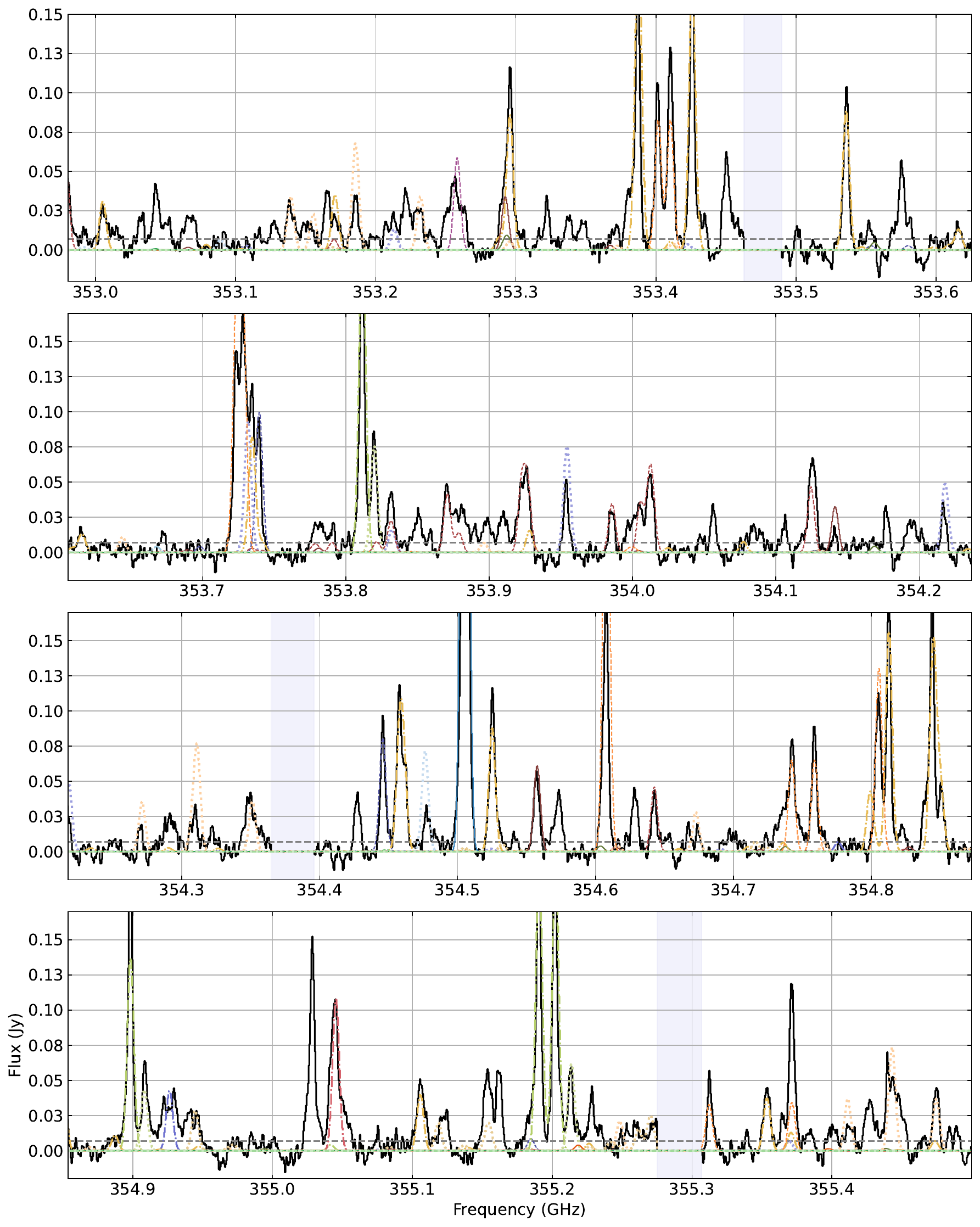}
    \caption{Continuation of figure \ref{spec}.\label{spec2}}
\end{figure}

\begin{figure}[!htbp]
    \centering
    \includegraphics[width=0.9\textwidth]{figures/molecules_legend.pdf} \\
    \includegraphics[width=0.9\textwidth]{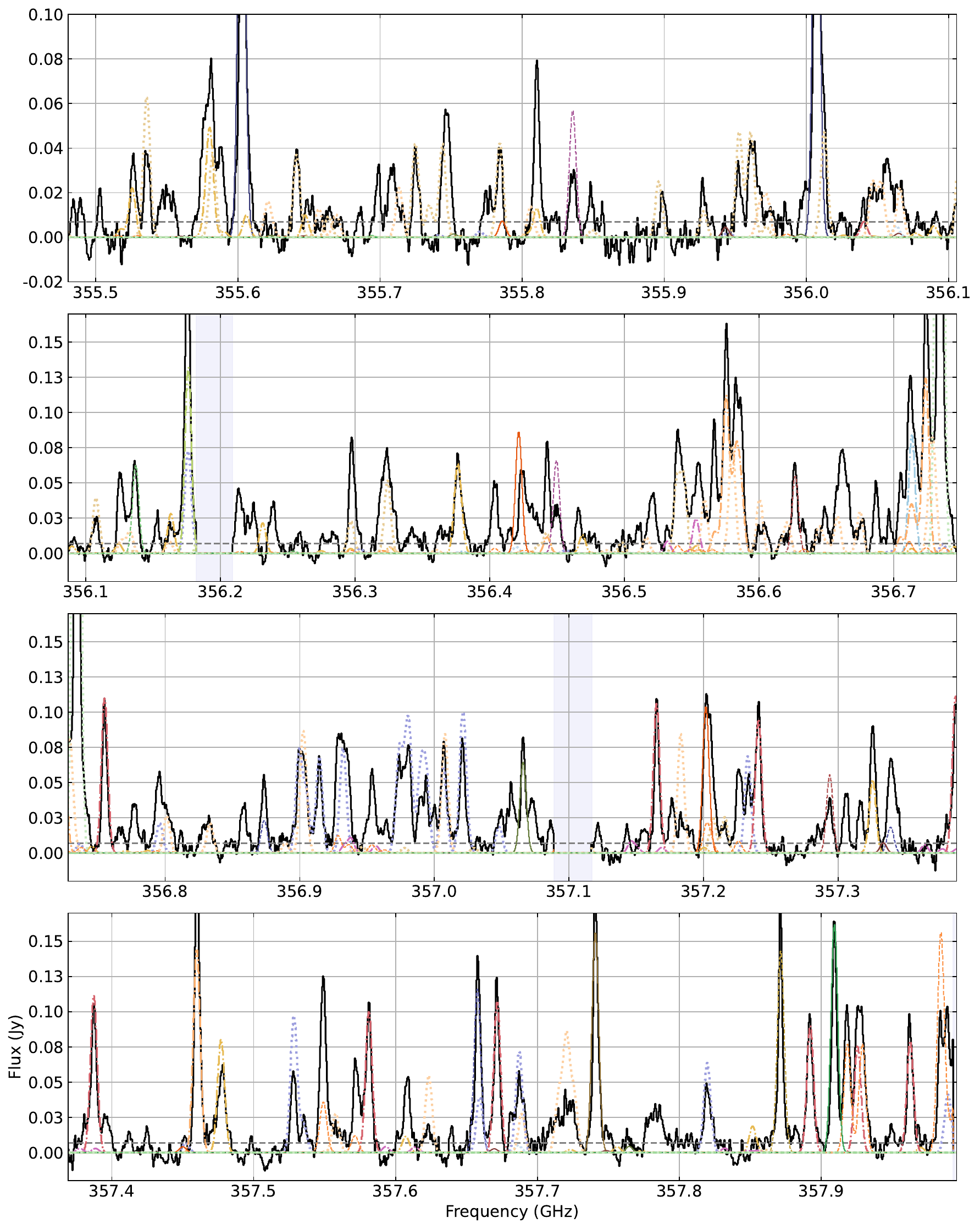}
    \caption{Continuation of figure \ref{spec}.\label{spec3}}
\end{figure}

\begin{figure}[!htbp]
    \centering
    \includegraphics[width=0.9\textwidth]{figures/molecules_legend.pdf} \\
    \includegraphics[width=0.9\textwidth]{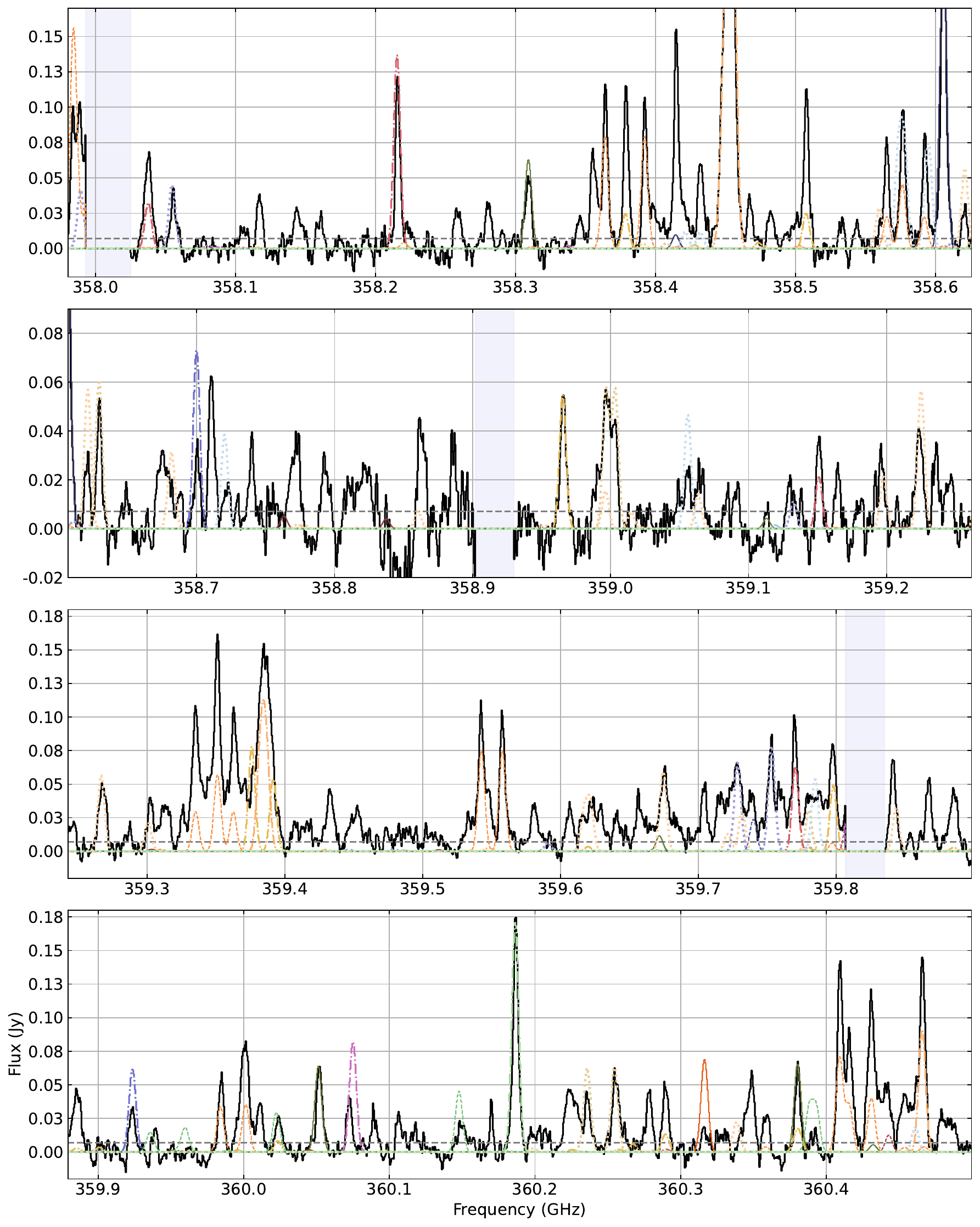}
    \caption{Continuation of figure \ref{spec}.\label{spec4}}
\end{figure}

\begin{figure}[!htbp]
    \centering
    \includegraphics[width=0.9\textwidth]{figures/molecules_legend.pdf} \\
    \includegraphics[width=0.9\textwidth]{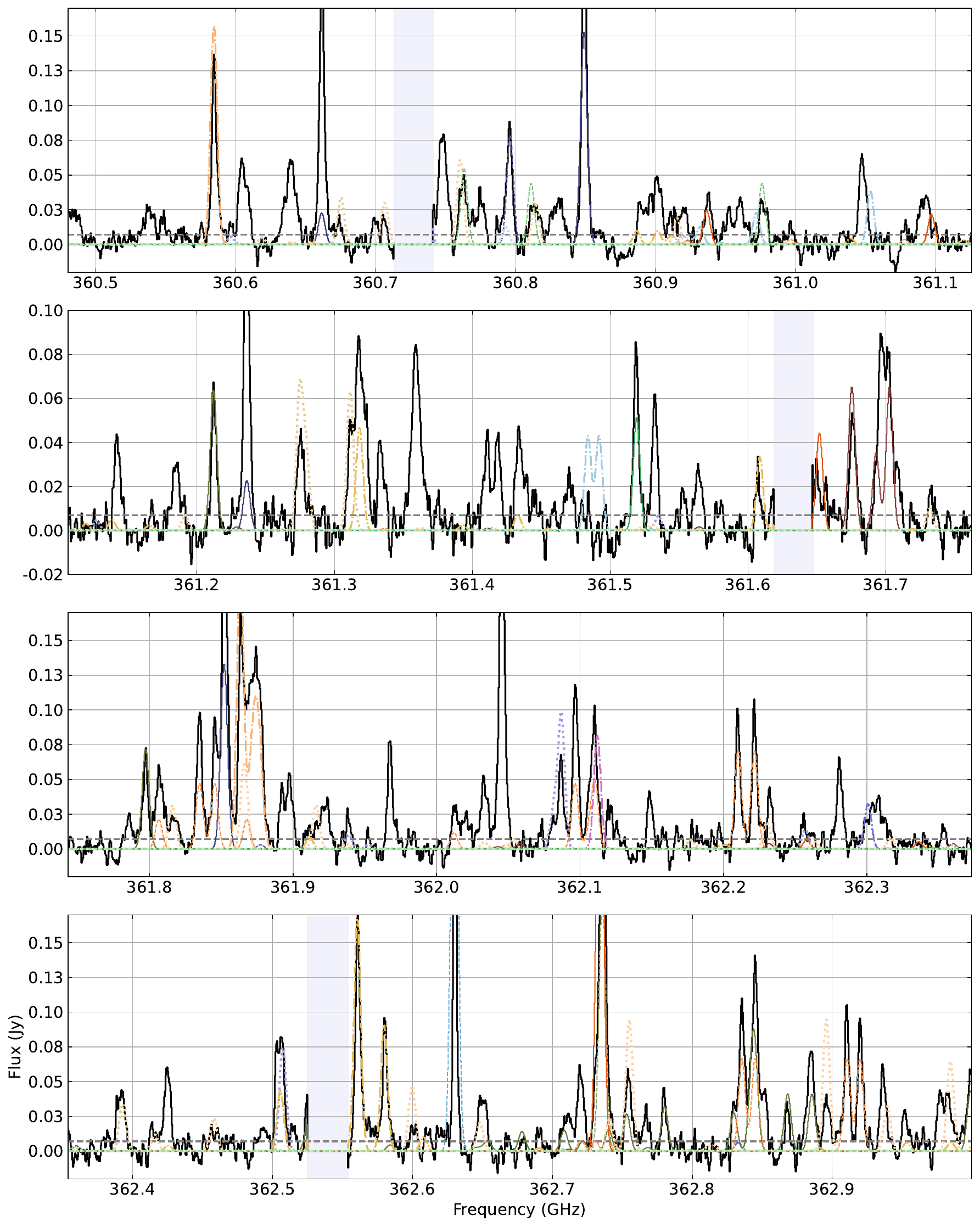}
    \caption{Continuation of figure \ref{spec}.\label{spec5}}
\end{figure}

\begin{figure}[!htbp]
    \centering
    \includegraphics[width=0.9\textwidth]{figures/molecules_legend.pdf} \\
    \includegraphics[width=0.9\textwidth]{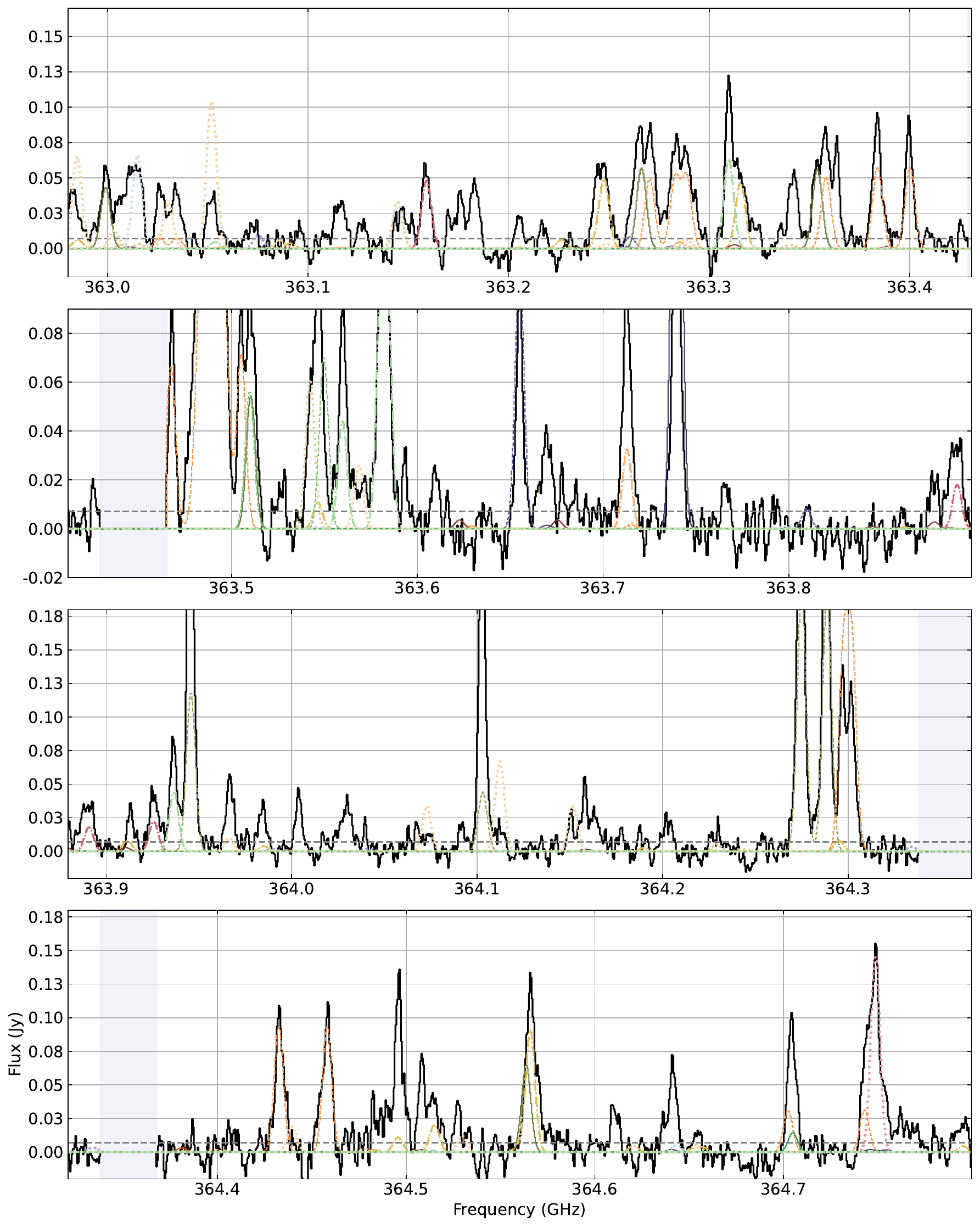}
    \caption{Continuation of figure \ref{spec}.\label{spec6}}
\end{figure}

\begin{figure}[!htbp]
    \centering
    \includegraphics[width=0.9\textwidth]{figures/molecules_legend.pdf} \\
    \includegraphics[width=0.9\textwidth]{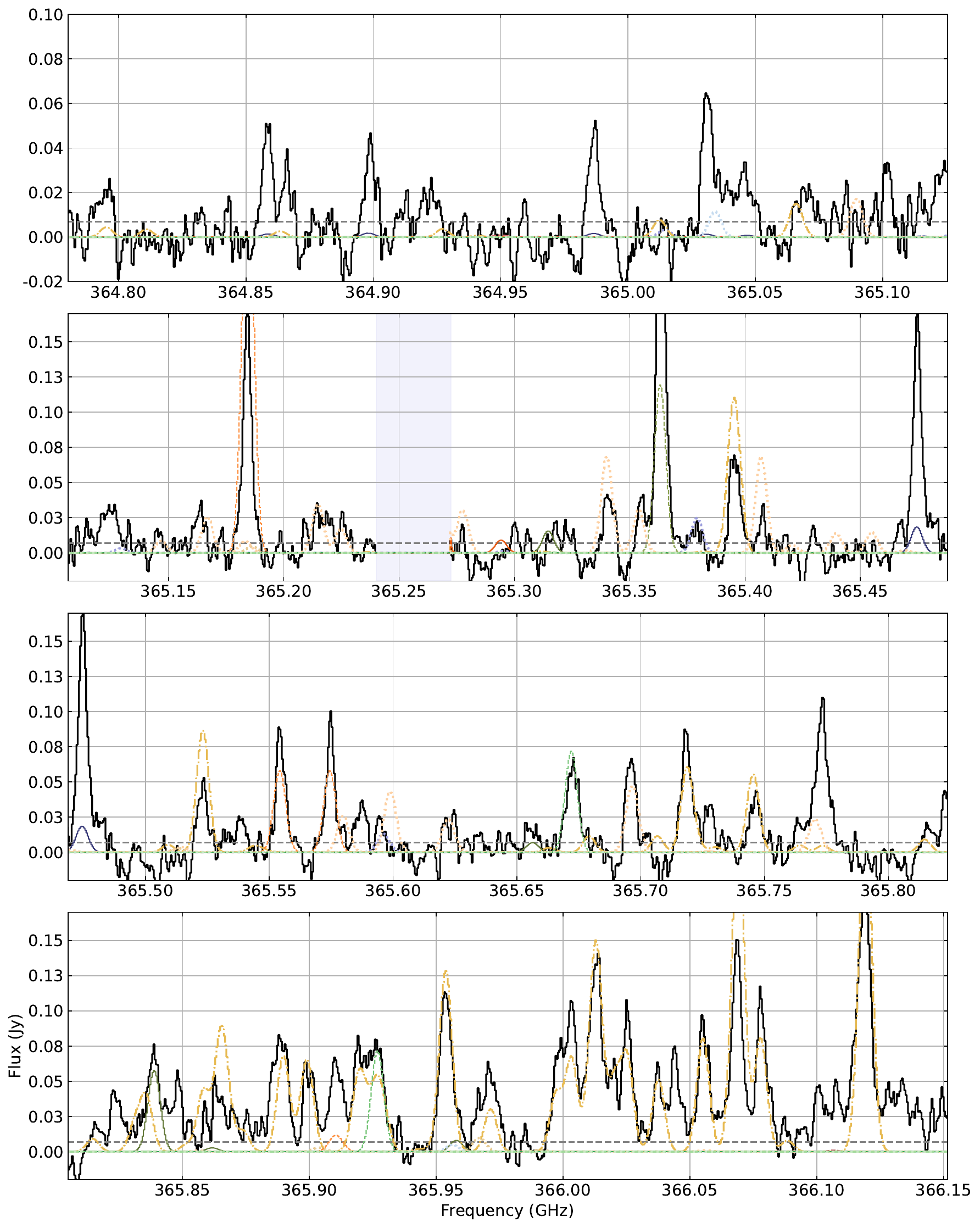}
    \caption{Continuation of figure \ref{spec}.\label{spec7}}
\end{figure}

\clearpage


\section{Appendix B}\label{AppendixB}

Figure~\ref{sumspec}, \ref{sumspec1}, \ref{sumspec2}, \ref{sumspec3}, \ref{sumspec4}, and \ref{sumspec5} represent summed spectra of the best-fit model of all detected molecules in blue, overlaid on top of the observed data in black. The horizontal gray dashed line represents the 1$\sigma$ noise level on the observed data.

\begin{figure}[!htbp]
    \centering
    \includegraphics[width=0.90\textwidth]{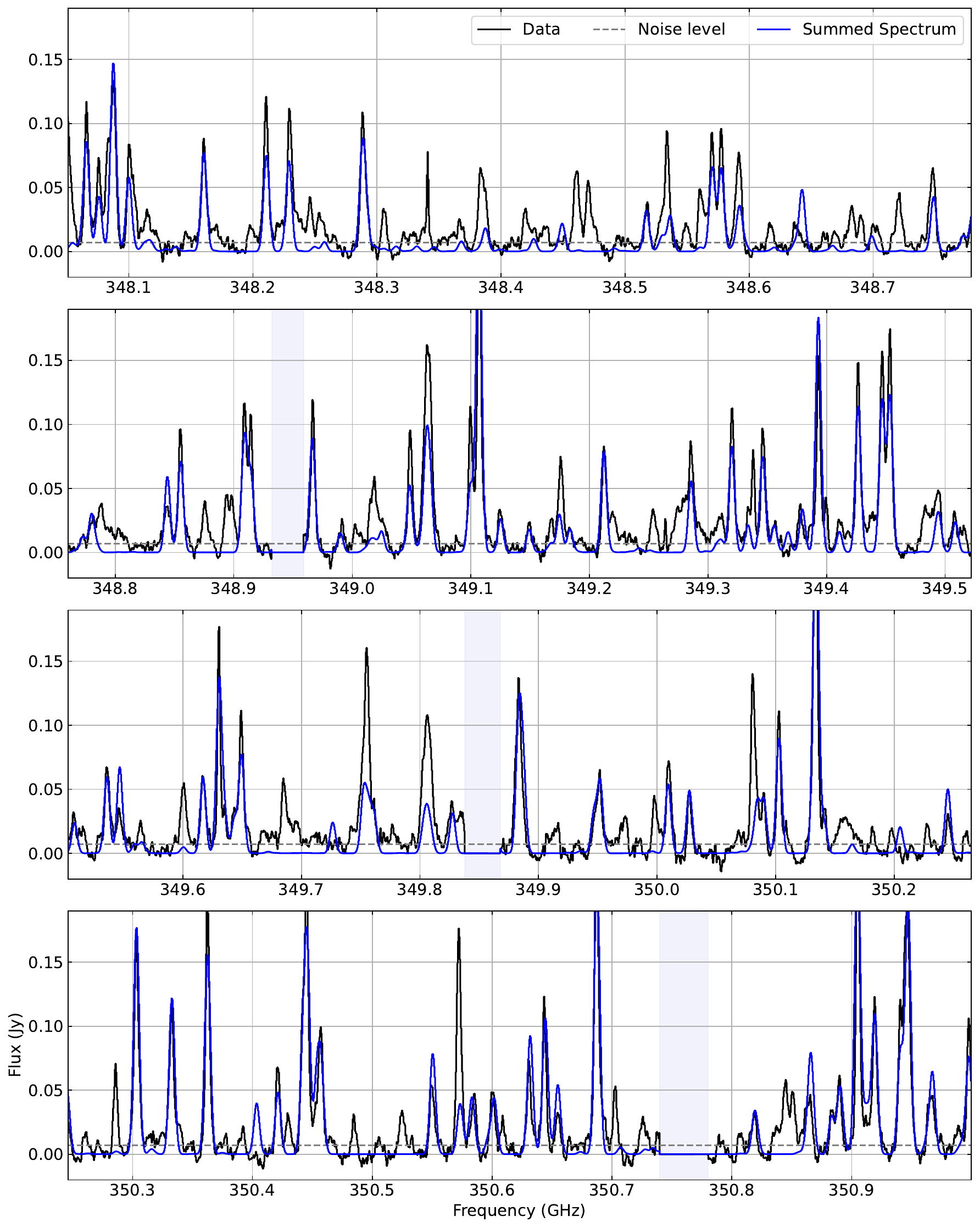}
    \caption{In black are the observed spectra, and in blue is the cumulative model adding the contribution from each molecule. The gaps between the spectral windows are excluded from the fit and shaded in lavender.\label{sumspec}}
\end{figure}

\begin{figure}[!htbp]
    \centering
    \includegraphics[width=0.90\textwidth]{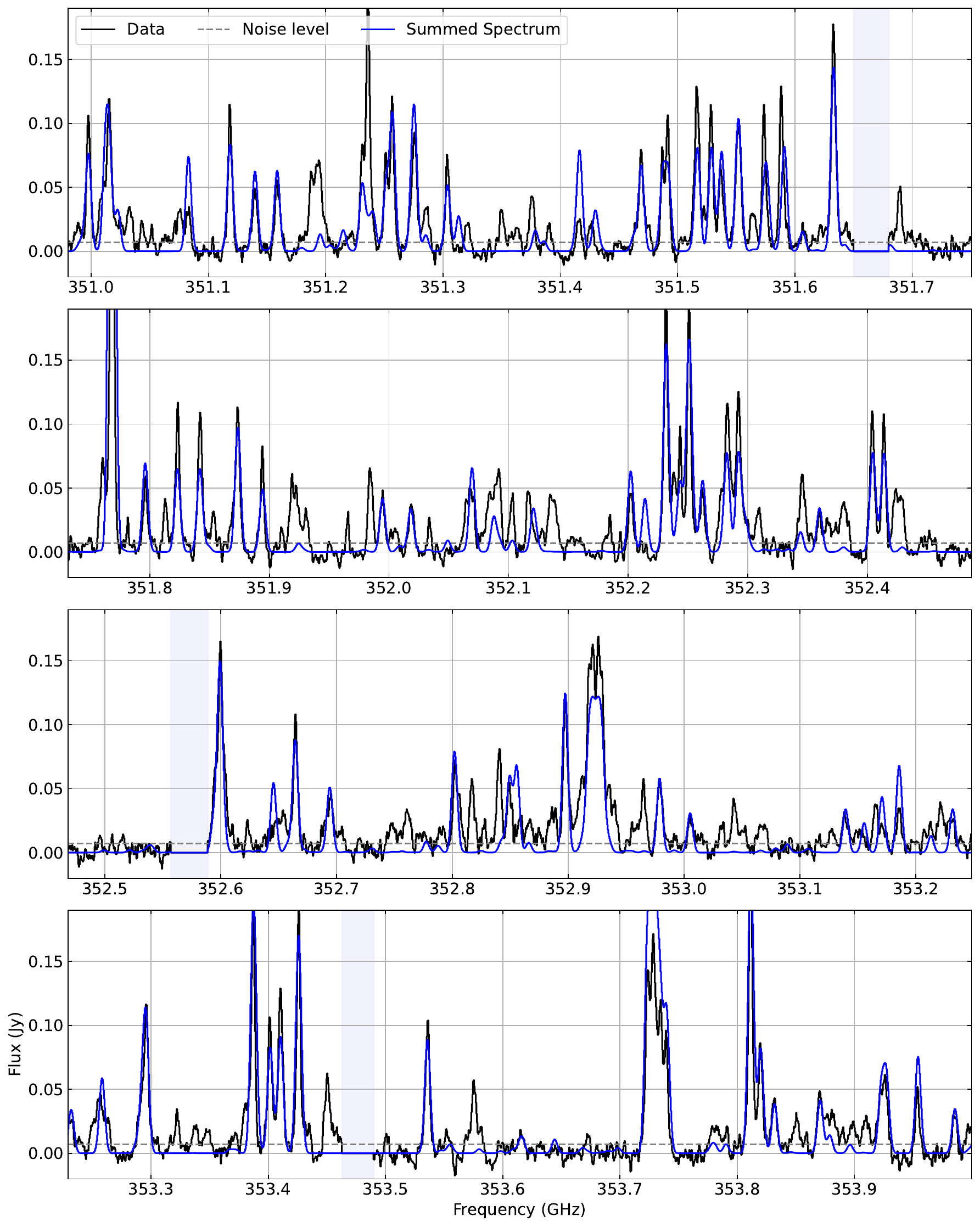}
    \caption{Continuation of figure \ref{sumspec}.\label{sumspec1}}
\end{figure}

\begin{figure}[!htbp]
    \centering
    \includegraphics[width=0.90\textwidth]{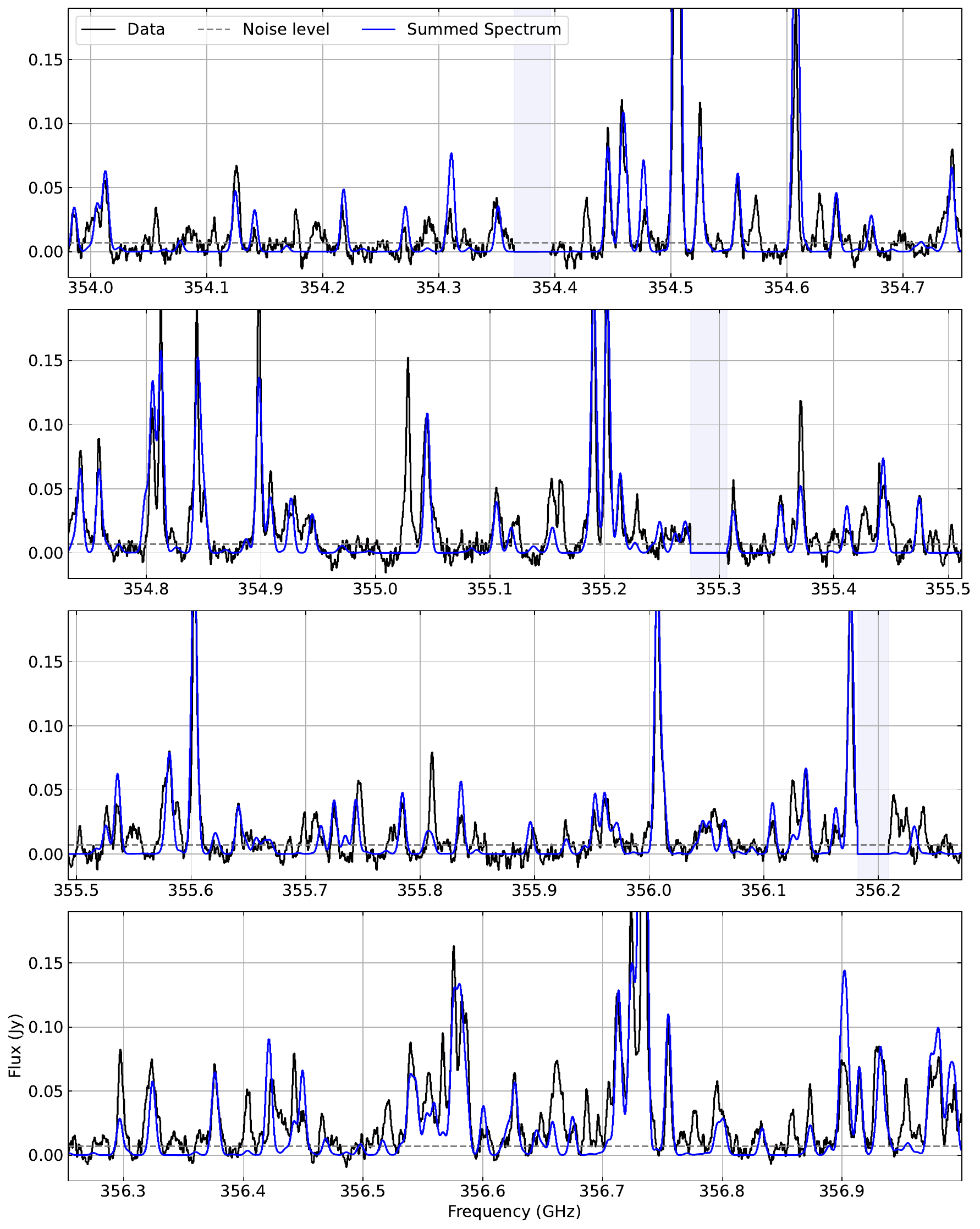}
    \caption{Continuation of figure \ref{sumspec}.\label{sumspec2}}
\end{figure}

\begin{figure}[!htbp]
    \centering
    \includegraphics[width=0.90\textwidth]{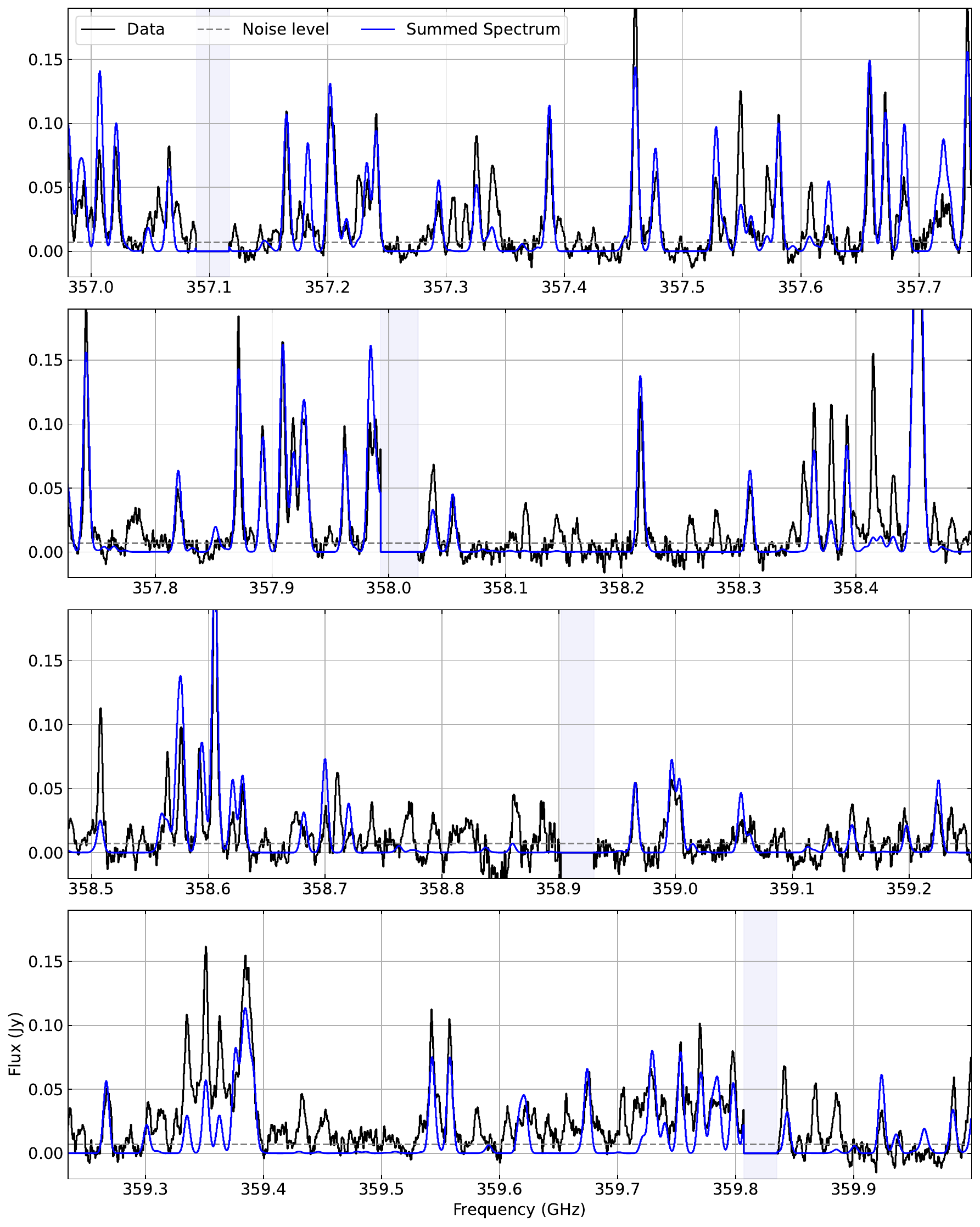}
    \caption{Continuation of figure \ref{sumspec}.\label{sumspec3}}
\end{figure}

\begin{figure}[!htbp]
    \centering
    \includegraphics[width=0.90\textwidth]{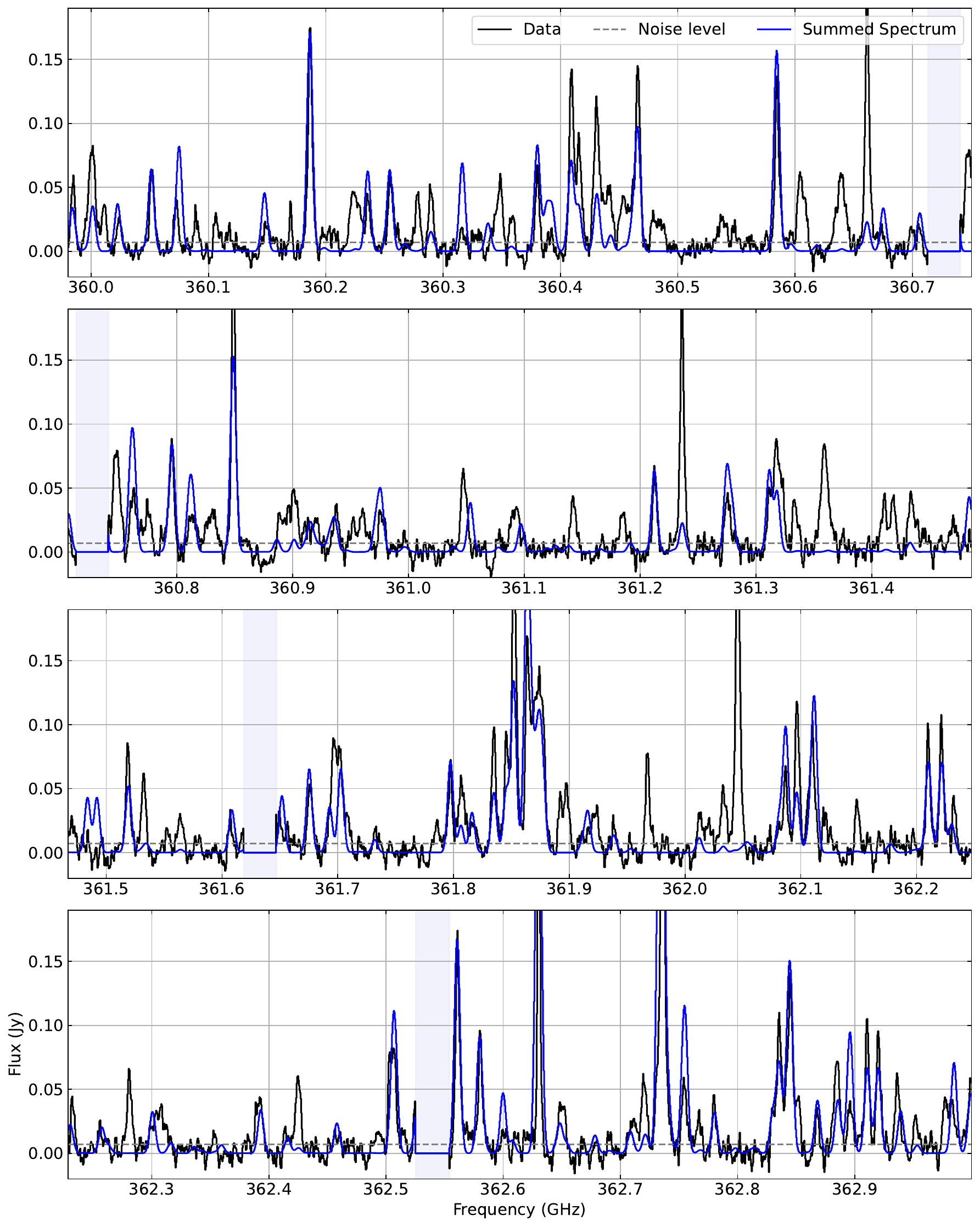}
    \caption{Continuation of figure \ref{sumspec}.\label{sumspec4}}
\end{figure}

\begin{figure}[!htbp]
    \centering
    \includegraphics[width=0.90\textwidth]{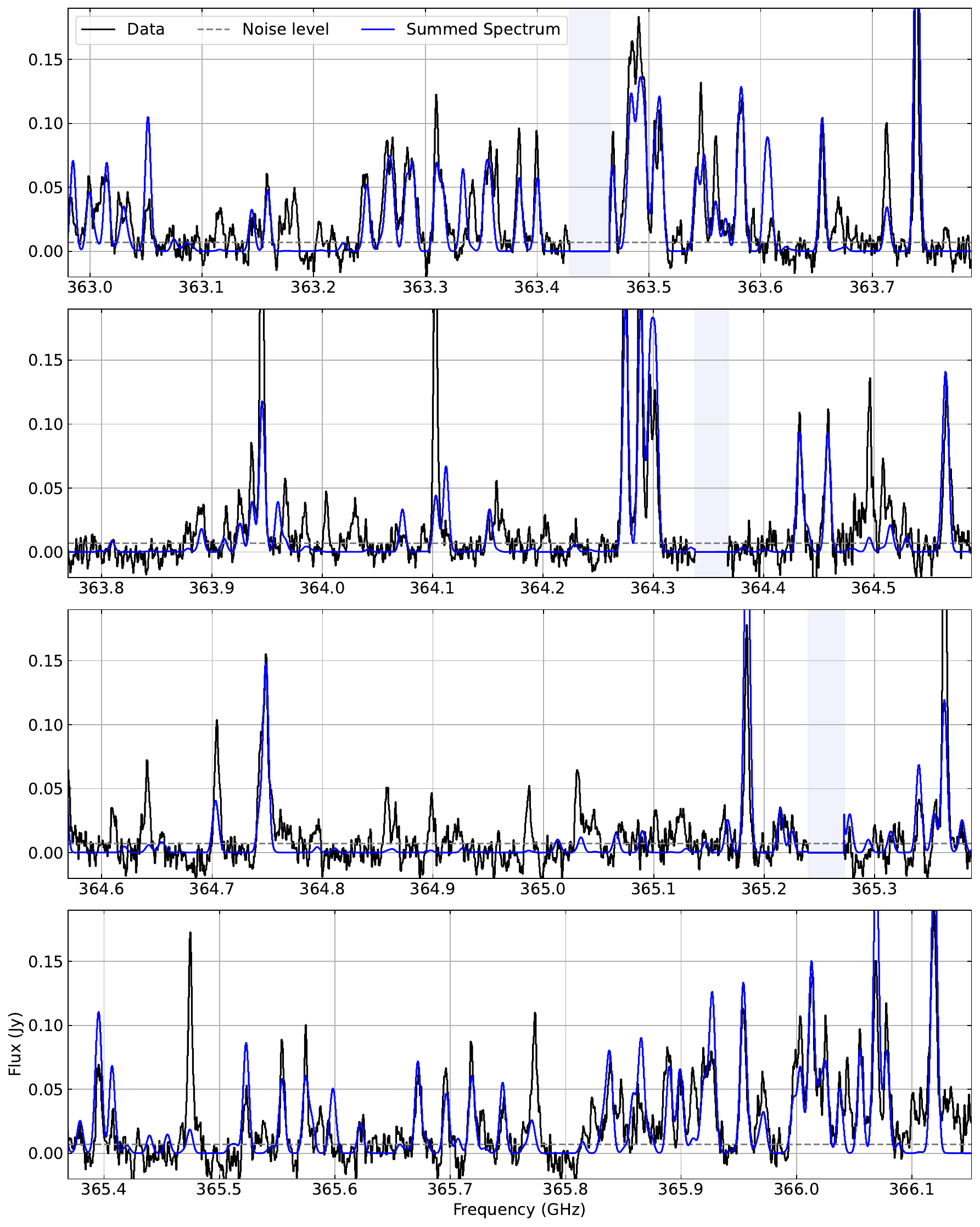}
    \caption{Continuation of figure \ref{sumspec}.\label{sumspec5}}
\end{figure}

\clearpage


\section{Appendix C}\label{AppendixC}

Figure~\ref{post}, \ref{post1}, \ref{post2}, and \ref{post3} represent the posterior distributions of the parameter space for the column density ($\mathrm{N_{tot}}$) and the excitation temperature ($\mathrm{T_{ex}}$) of the molecules, showing that these two parameters are well constrained by the model. We ran an MCMC with 100 walkers and 10,000 steps, discarding the first 2,000 steps as burn-in, and using the remaining 8,000 steps to demonstrate the posterior distribution.

\begin{figure}[!htbp]
\gridline{\fig{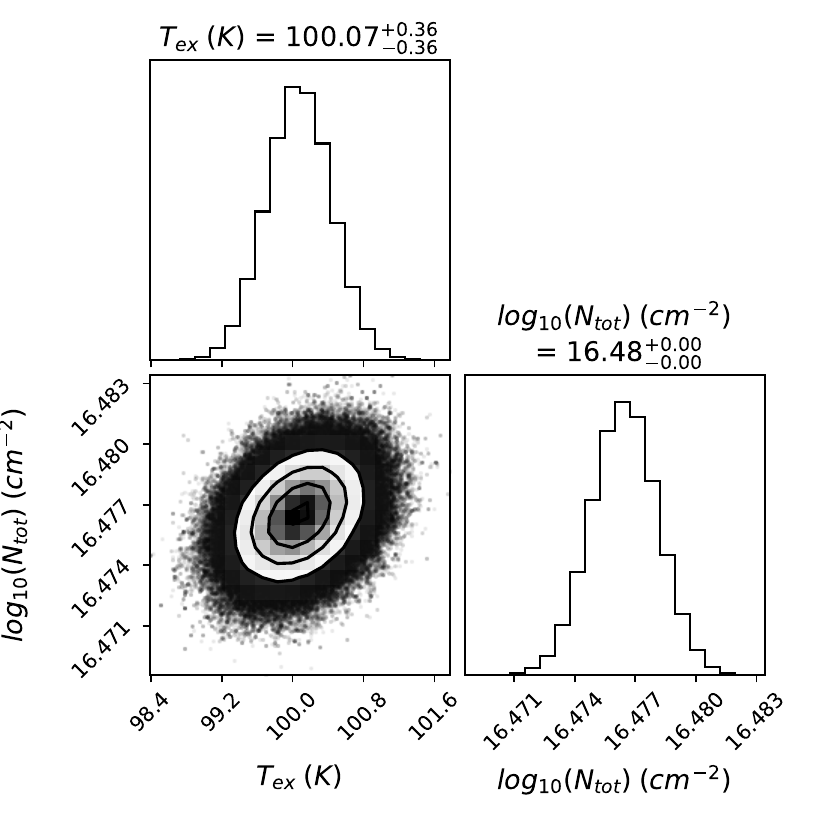}{0.34\textwidth}{Methanol $\mathrm{CH}_3\mathrm{OH}$}
          \fig{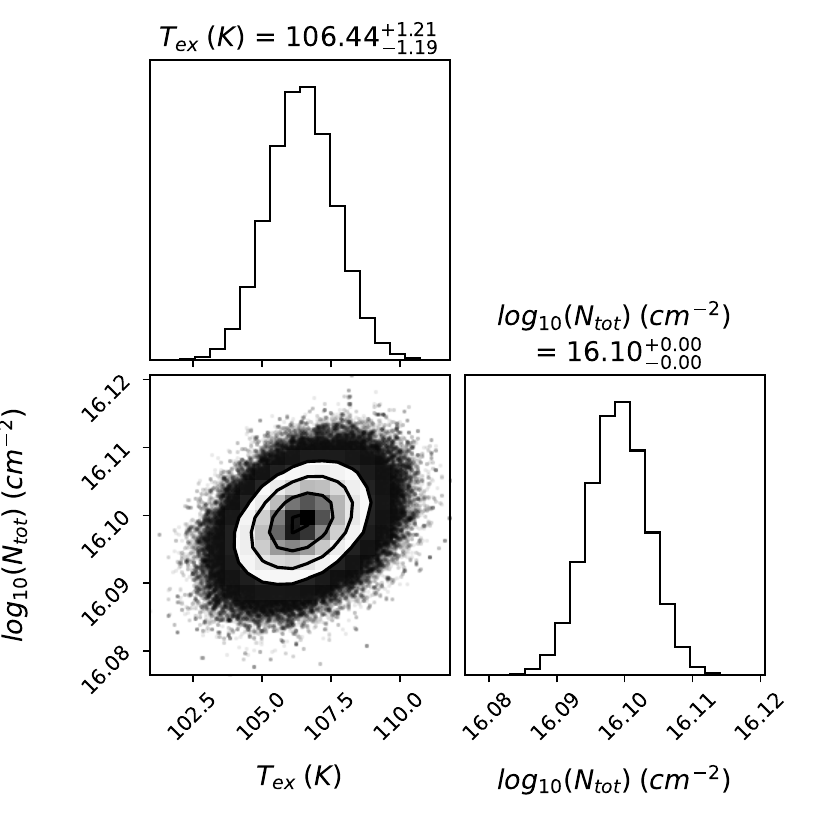}{0.34\textwidth}{Methanol $\mathrm{^{13}CH}_3\mathrm{OH}$}}
\gridline{\fig{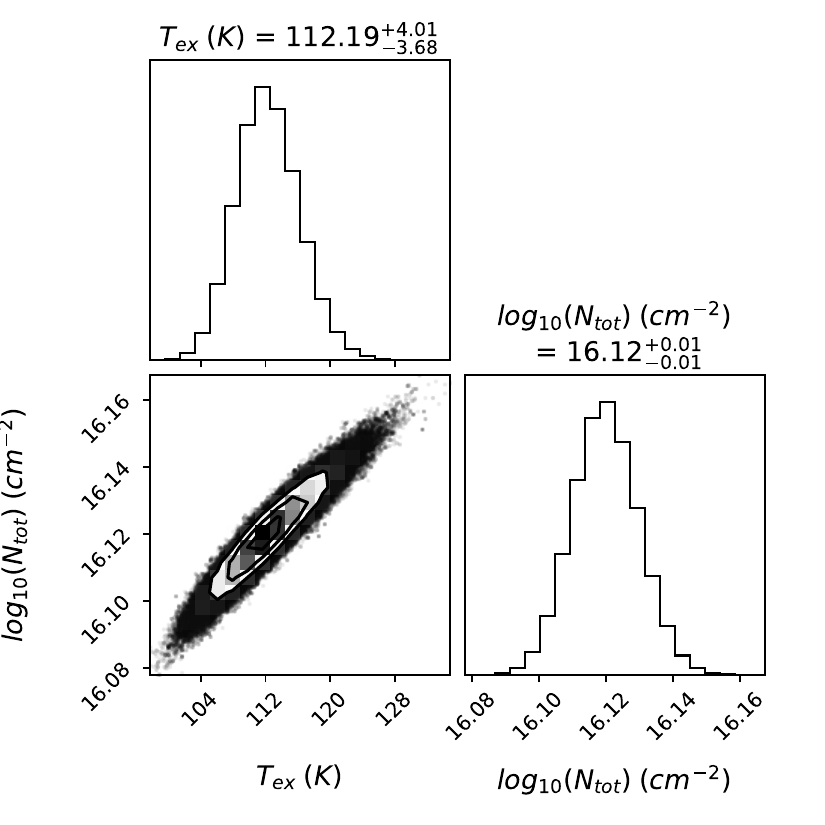}{0.34\textwidth}{Methanol $\mathrm{CH}_2\mathrm{DOH}$}
          \fig{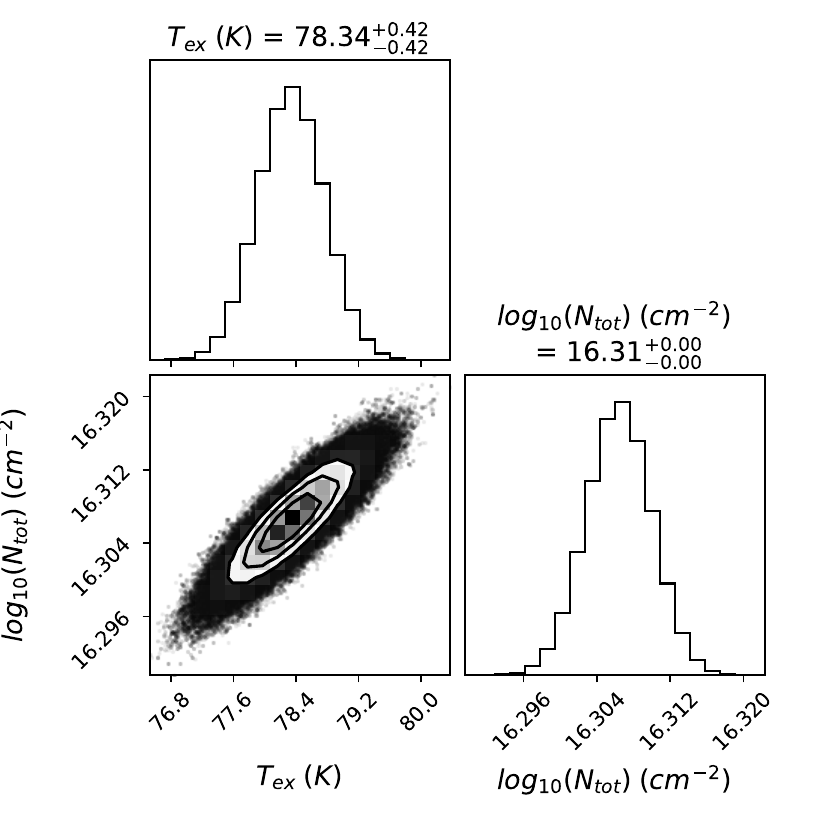}{0.34\textwidth}{Dimethyl ether $\mathrm{CH}_3\mathrm{OCH}_3$}}
\gridline{\fig{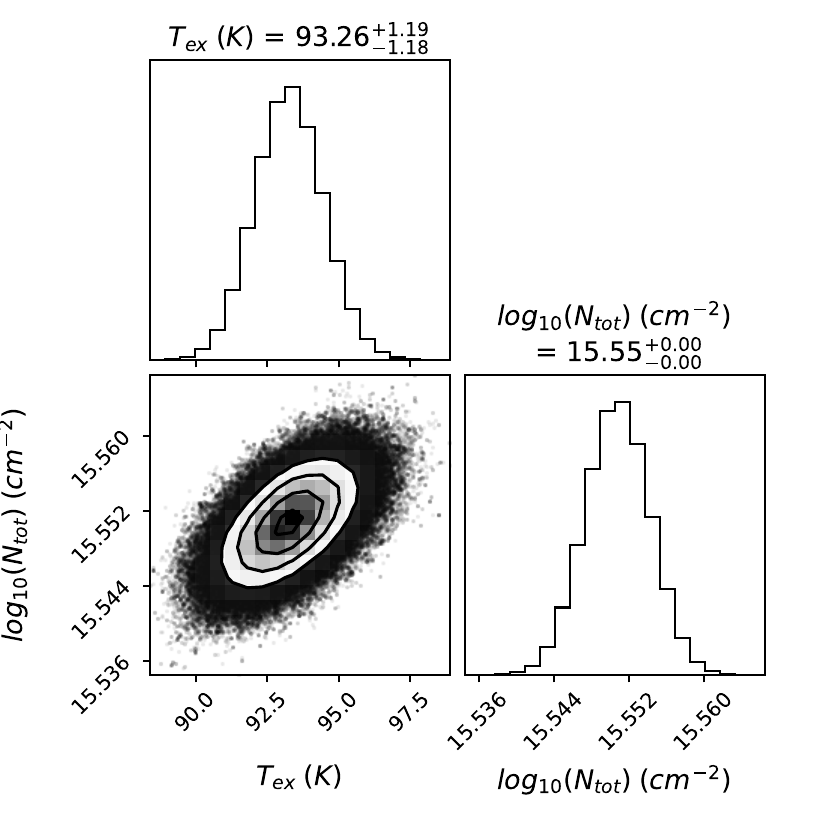}{0.34\textwidth}{ Sulfur dioxide $\mathrm{SO}_2$}
          \fig{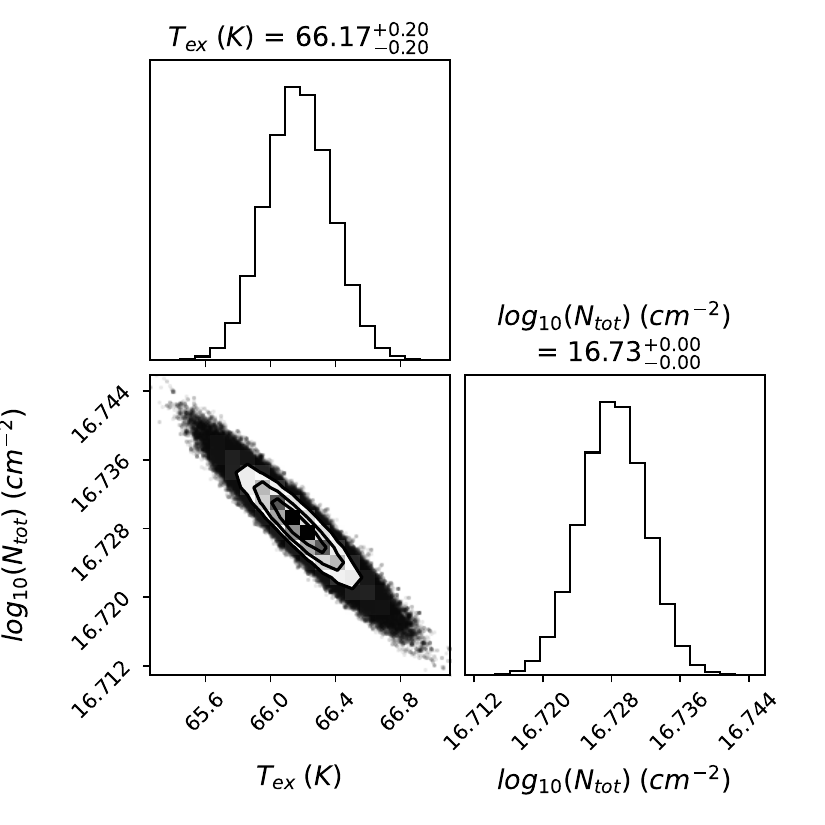}{0.34\textwidth}{Methyl Formate $\mathrm{CH}_3\mathrm{OCHO}$}}
\caption{Posterior distributions for the excitaion temperature $T_{ex}$ and the column density $N_{tot}$ for various COMs.\label{post}}
\end{figure}

\begin{figure}[!htbp]
\gridline{\fig{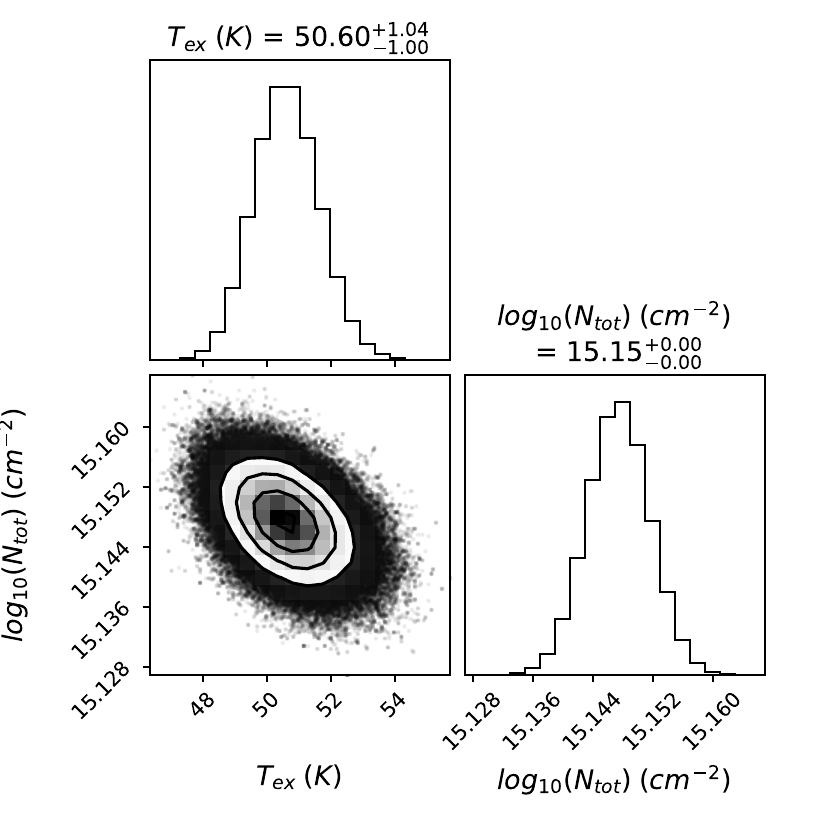}{0.34\textwidth}{Ethylene Oxide $\mathrm{\text{c-}H}_2\mathrm{COCH}_2$}
          \fig{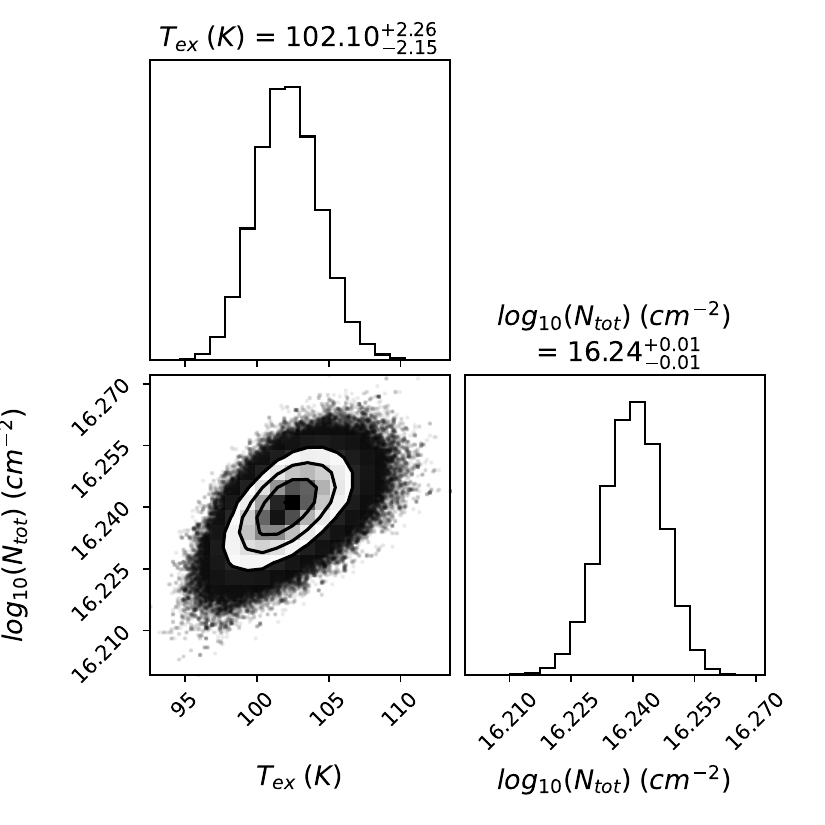}{0.34\textwidth}{ Acetone $\mathrm{CH}_3\mathrm{COCH}_3$}}
\gridline{\fig{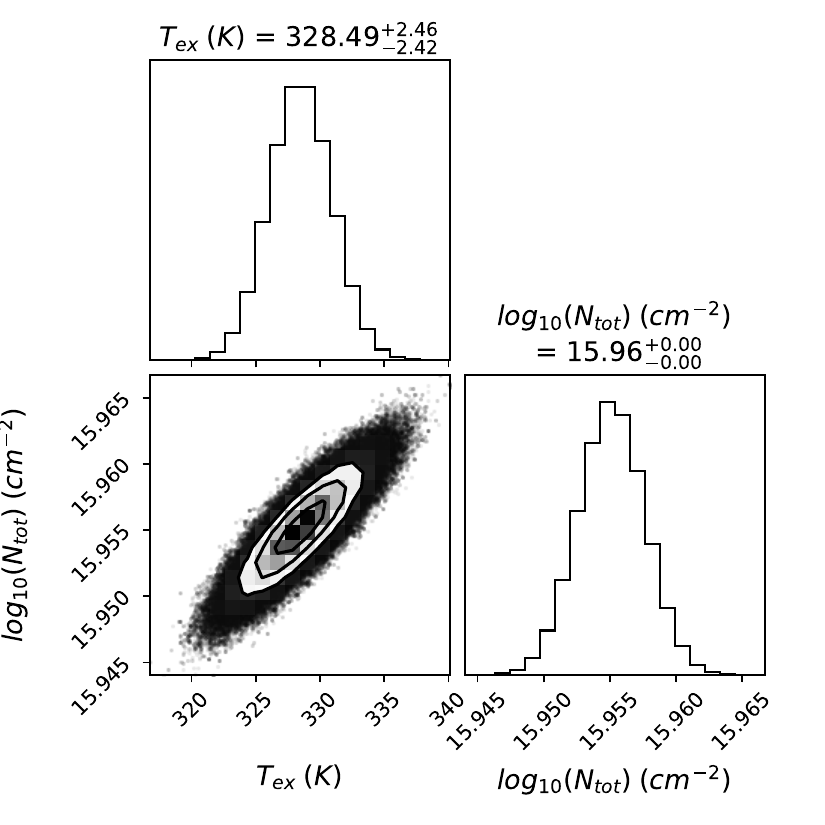}{0.34\textwidth}{Acetaldehyde $\mathrm{CH}_3\mathrm{CHO}$}
          \fig{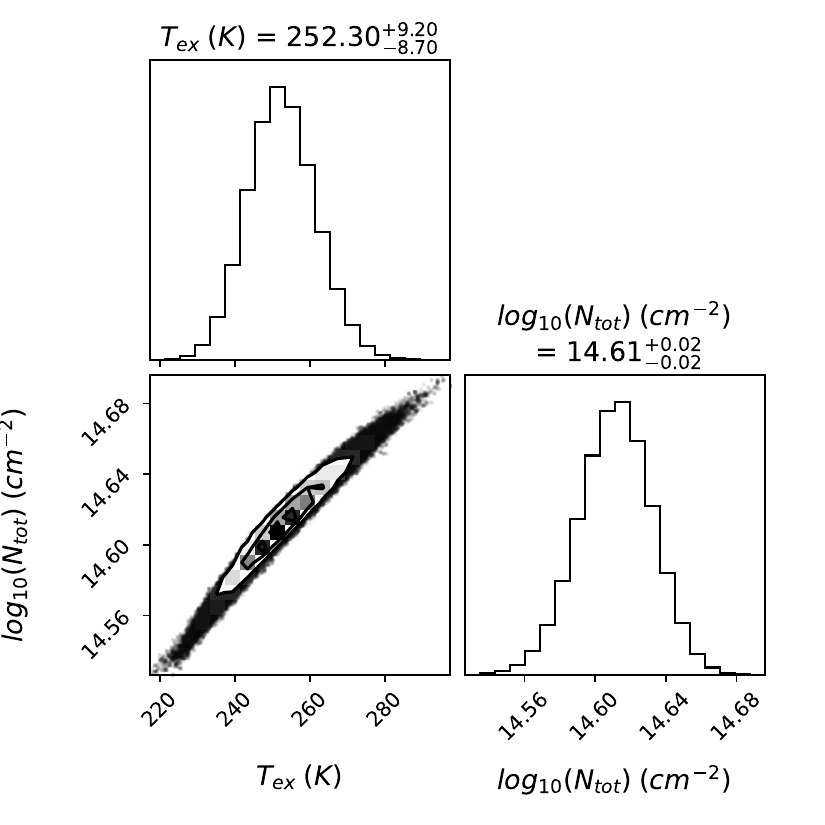}{0.34\textwidth}{Acetonitrile $\mathrm{CH}_3\mathrm{CN}$}}
\gridline{\fig{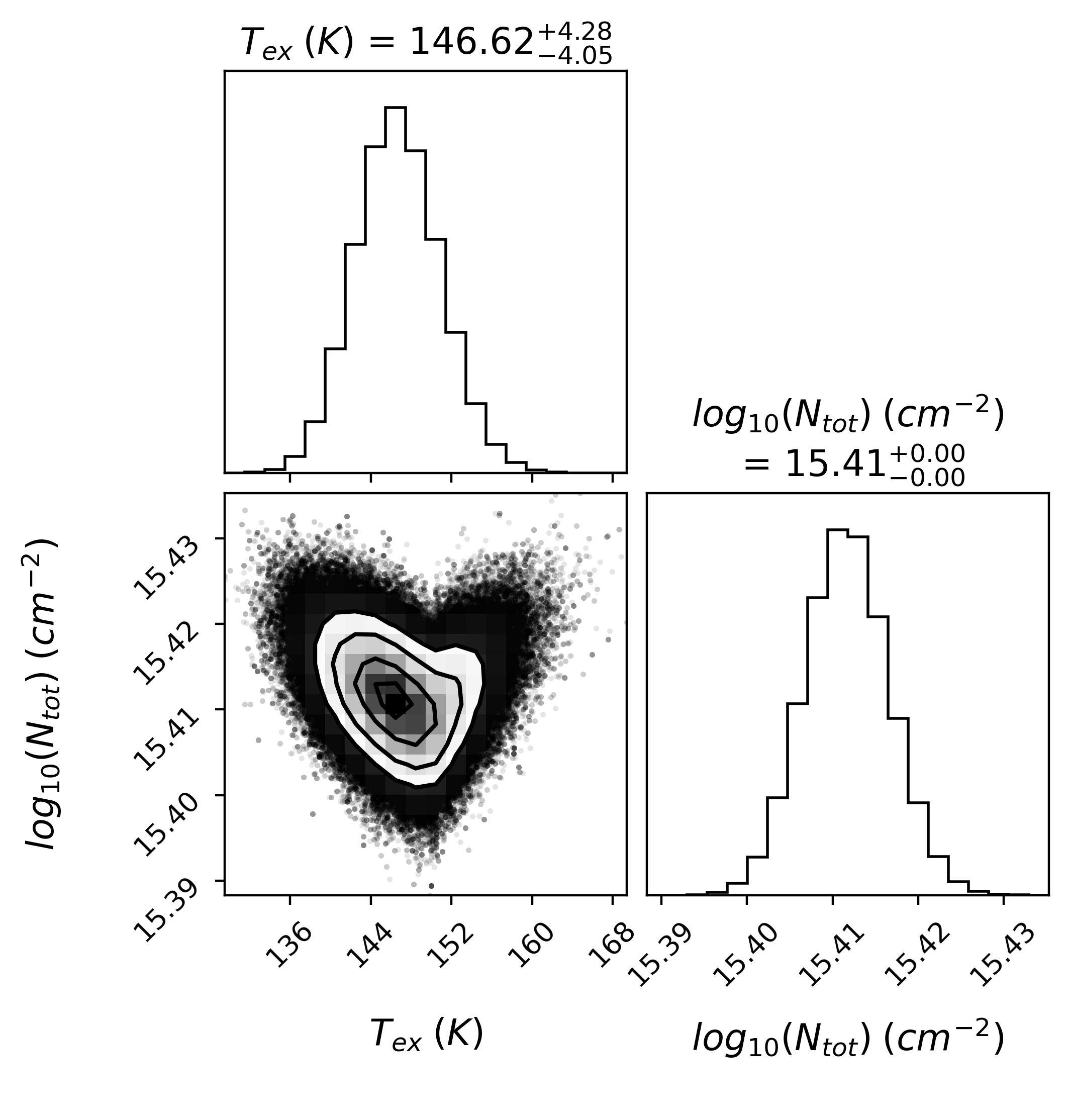}{0.34\textwidth}{ Ketene $\mathrm{H_{2}CCO}$}
          \fig{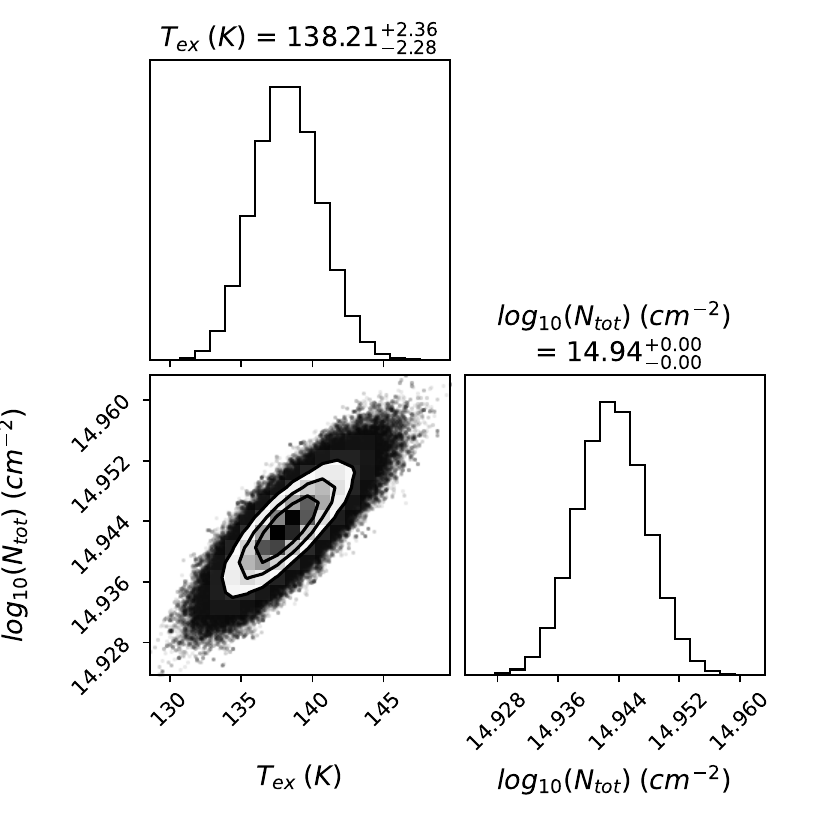}{0.34\textwidth}{Formaldehyde  $\mathrm{H}_2\mathrm{^{13}CO}$}}
\caption{Continuation of figure \ref{post}.\label{post1}}
\end{figure}

\begin{figure}[!htbp]
\gridline{\fig{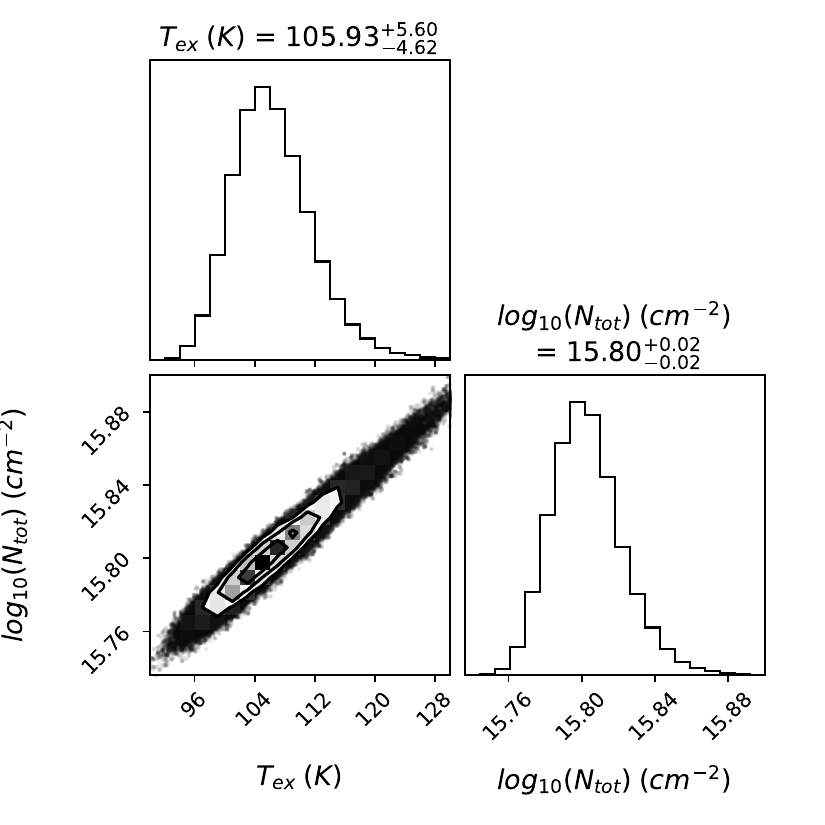}{0.34\textwidth}{Methanol $\mathrm{CH_{3}OD}$}
        \fig{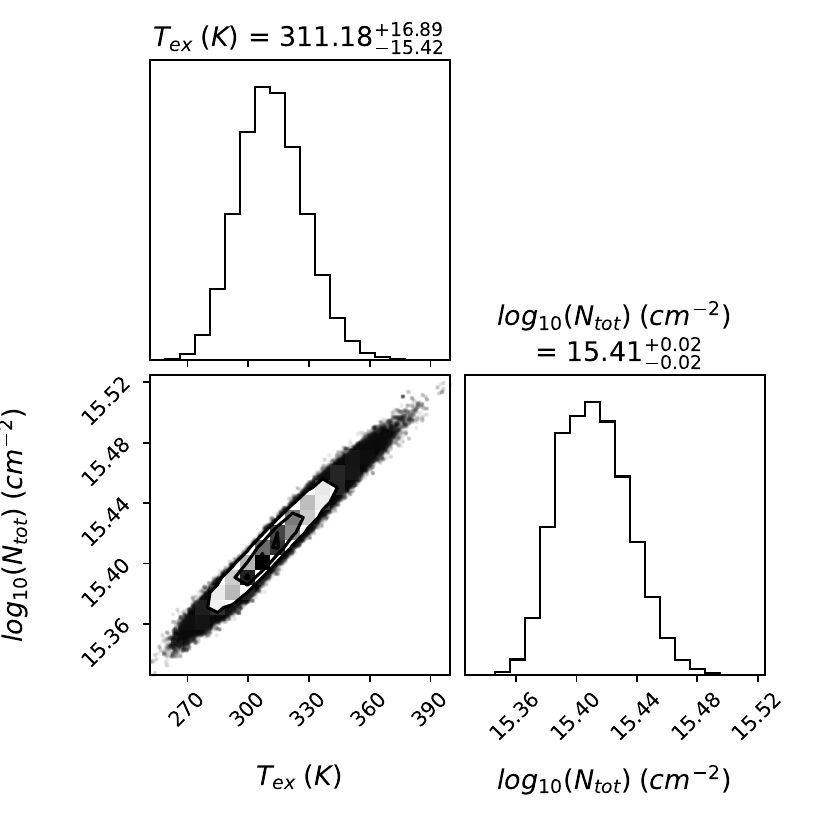}{0.34\textwidth}{Isocyanic Acid $\mathrm{HNCO}$}}
\gridline{\fig{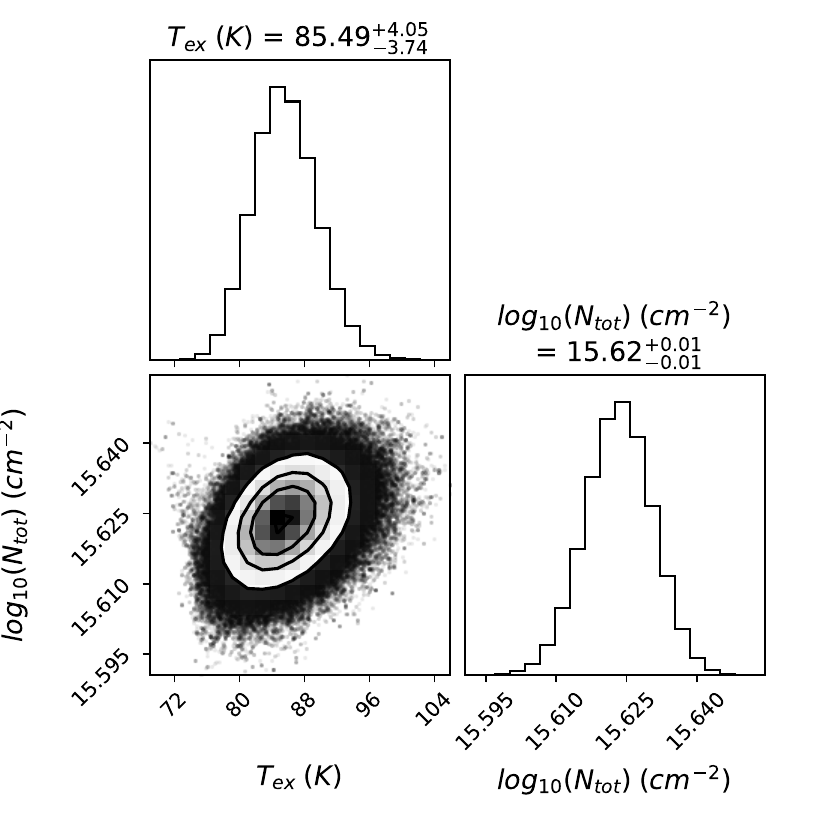}{0.34\textwidth}{Methyl Mercaptan $\mathrm{CH_3SH}$}
        \fig{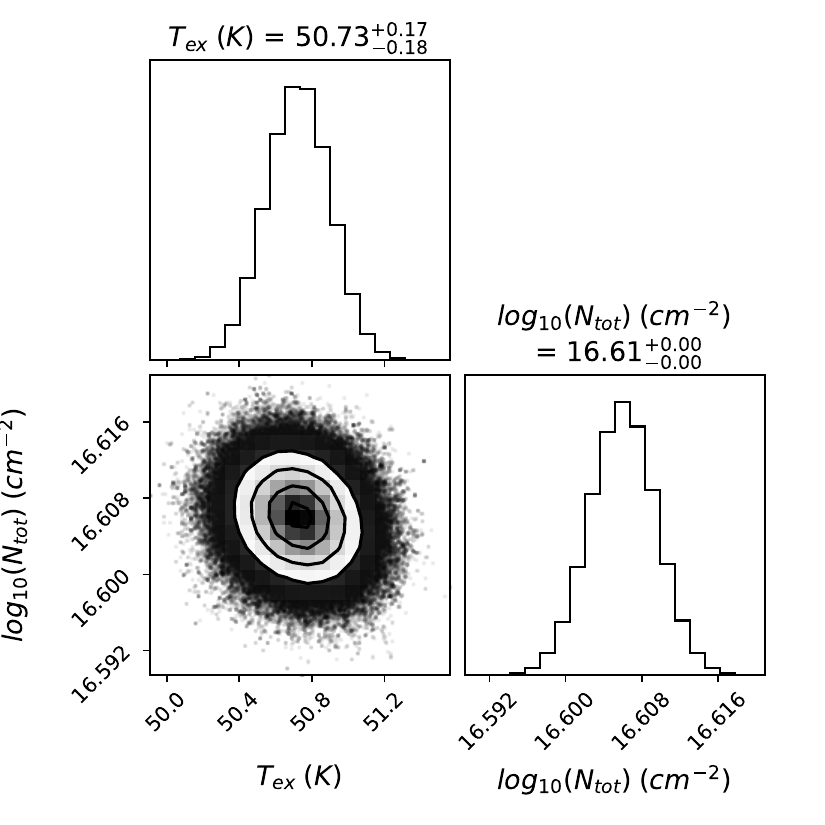}{0.34\textwidth}{Vinyl Cyanide $\mathrm{CH_2CHCN}$}}
\gridline{\fig{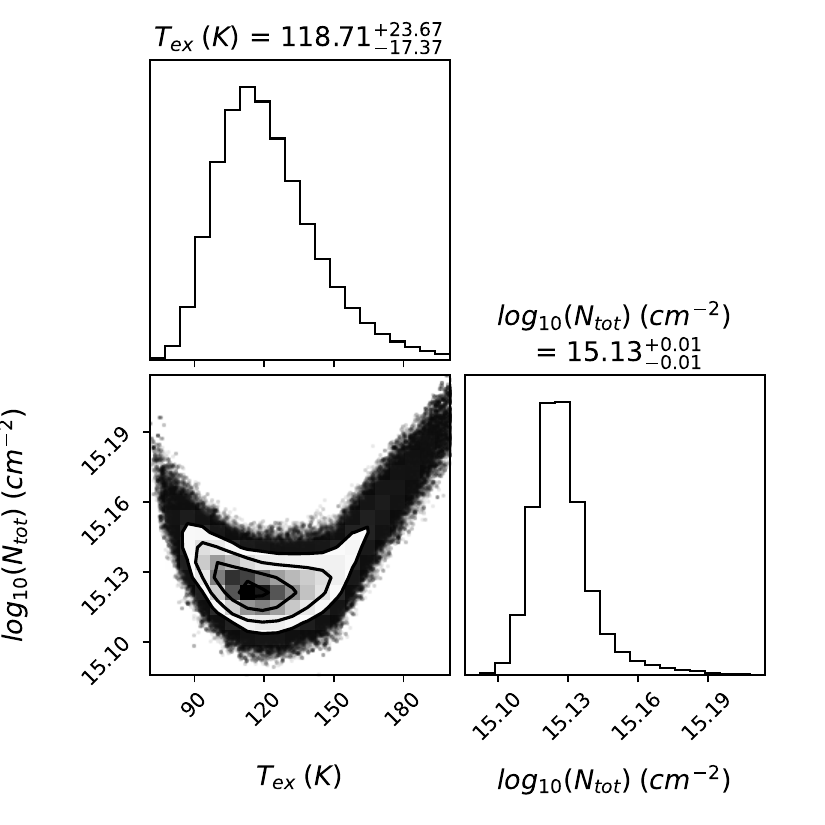}{0.34\textwidth}{Formic Acid $\mathrm{\text{t-}HCOOH}$}
        \fig{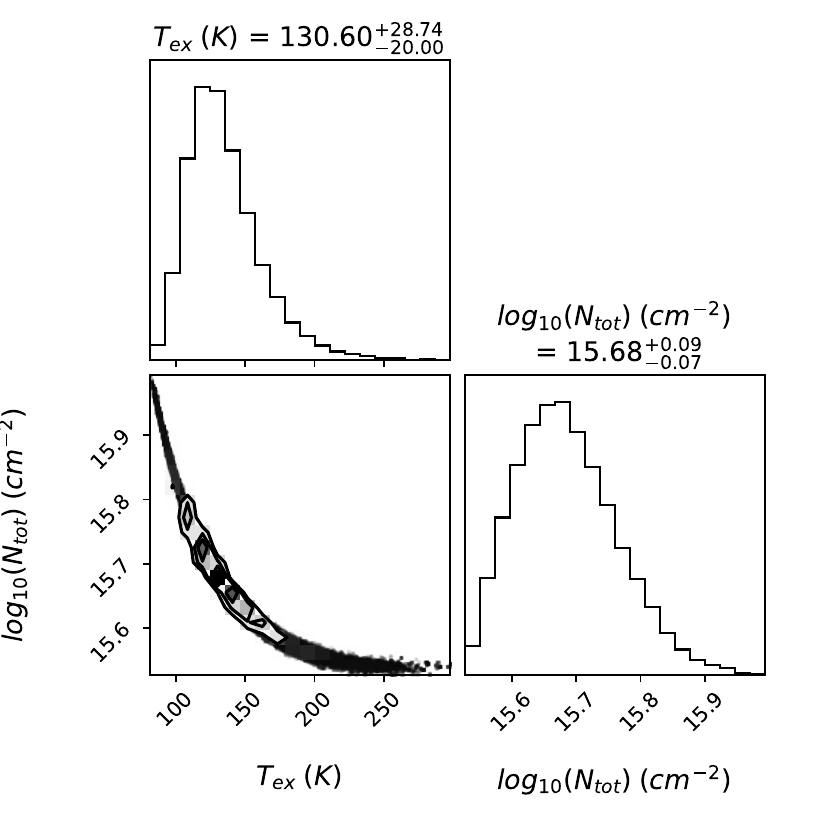}{0.34\textwidth}{Carbonyl Sulfide $\mathrm{OCS}$}}
\caption{Continuation of figure \ref{post}.\label{post2}}
\end{figure}

\clearpage

\begin{figure}[!htbp]
\gridline{\fig{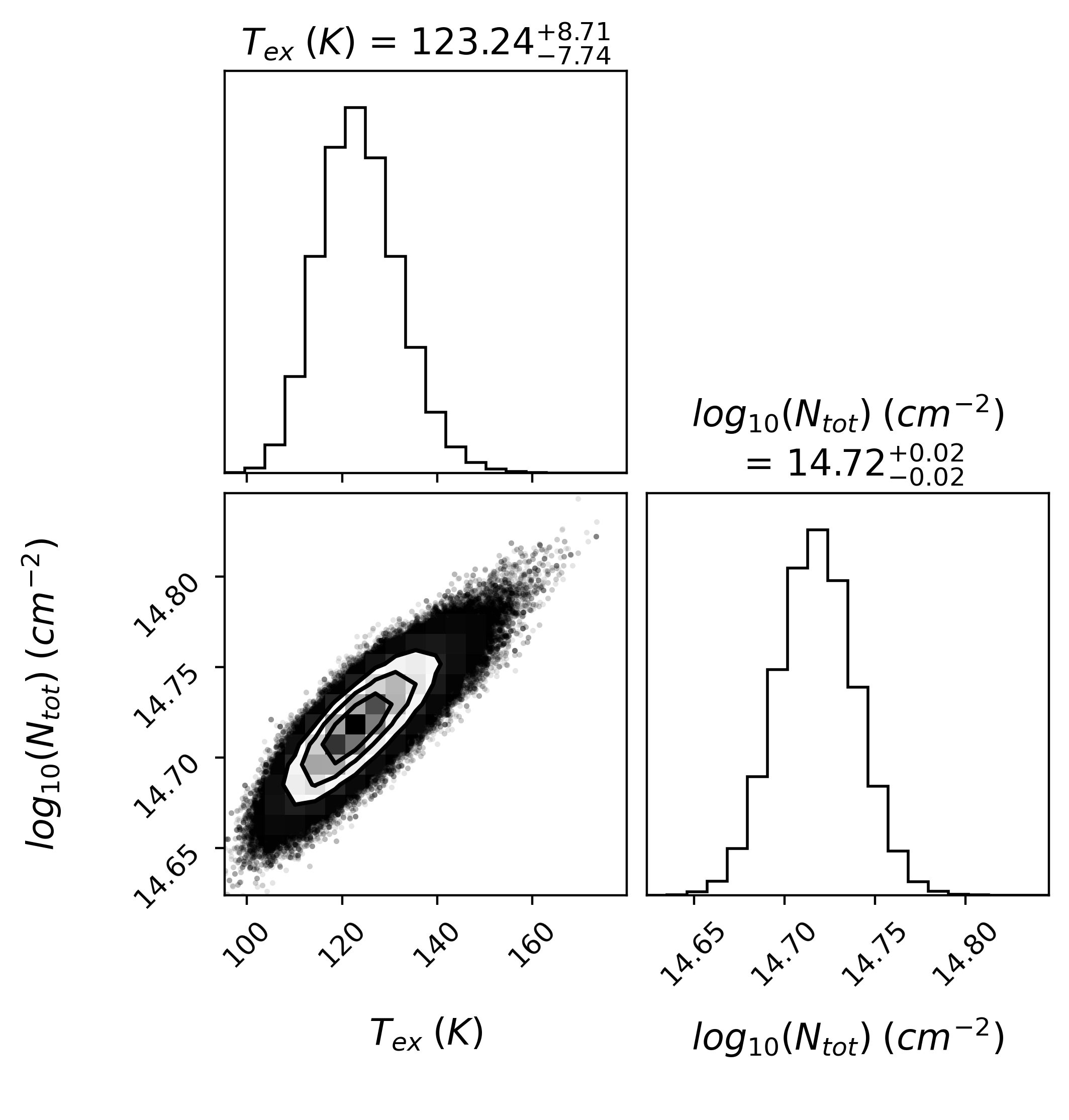}{0.80\textwidth}{Formaldehyde $\mathrm{H_2C^{17}O}$}}
\caption{Continuation of figure \ref{post}.\label{post3}}
\end{figure}

\clearpage


\section{Appendix D}\label{AppendixD}\subsection{Glycolaldehyde plots}

Figure~\ref{glad} and \ref{glad1} show the predicted Glycolaldehyde spectra by our model in blue, plotted on top of the observed data in black. The horizontal dashed line in gray represents the noise level in the data ($1\sigma$).

\begin{figure}[!htbp]
\gridline{\fig{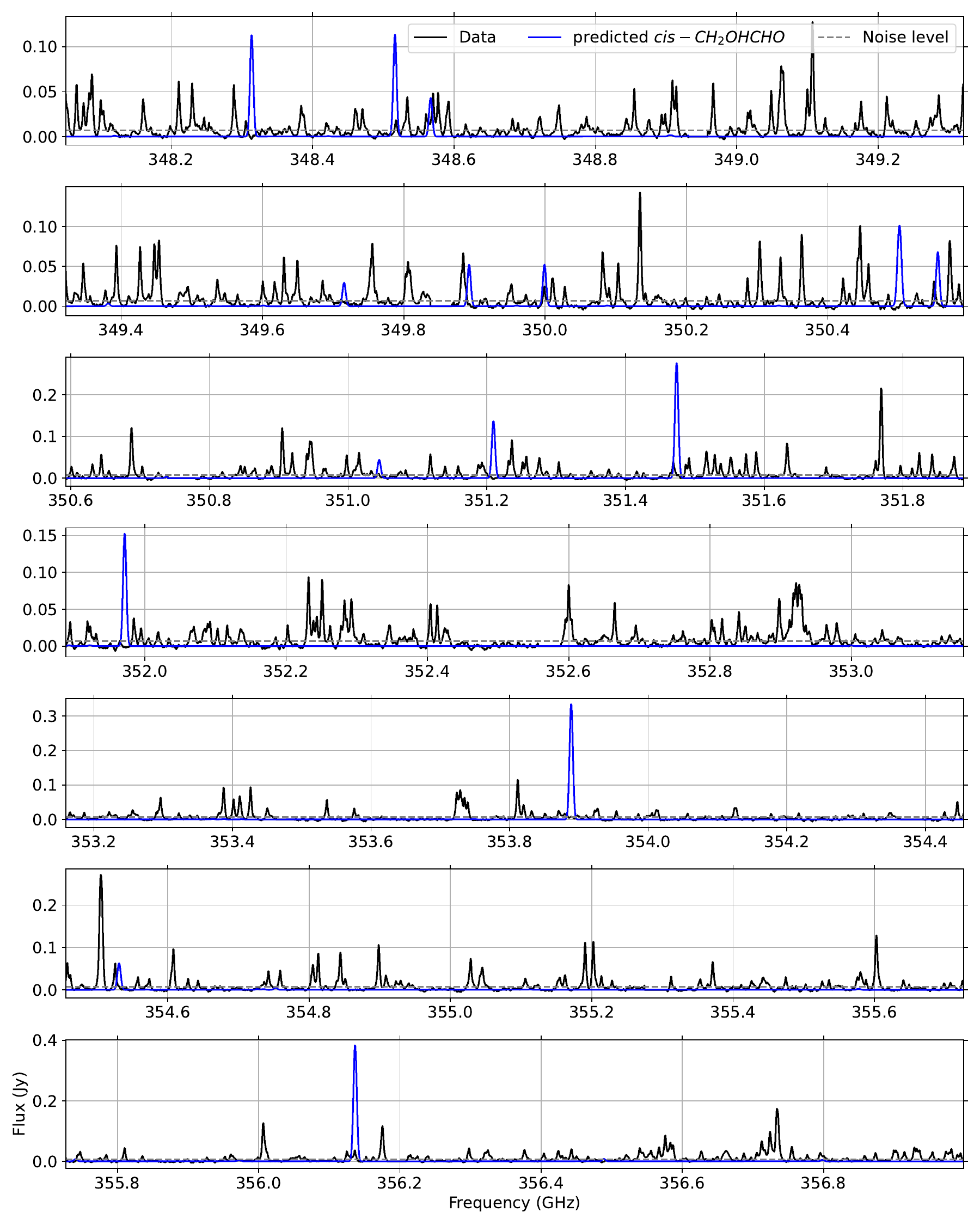}{0.87\textwidth}{}}
\caption{Predicted emission of the molecule Glycolaldehyde at $T_{ex}$ = 150 K and $N_{tot}$ = $4.0 \times 10 ^{16}\; \mathrm{cm^{-2}}$.}
\label{glad}
\end{figure}

\begin{figure}[!htbp]
\gridline{\fig{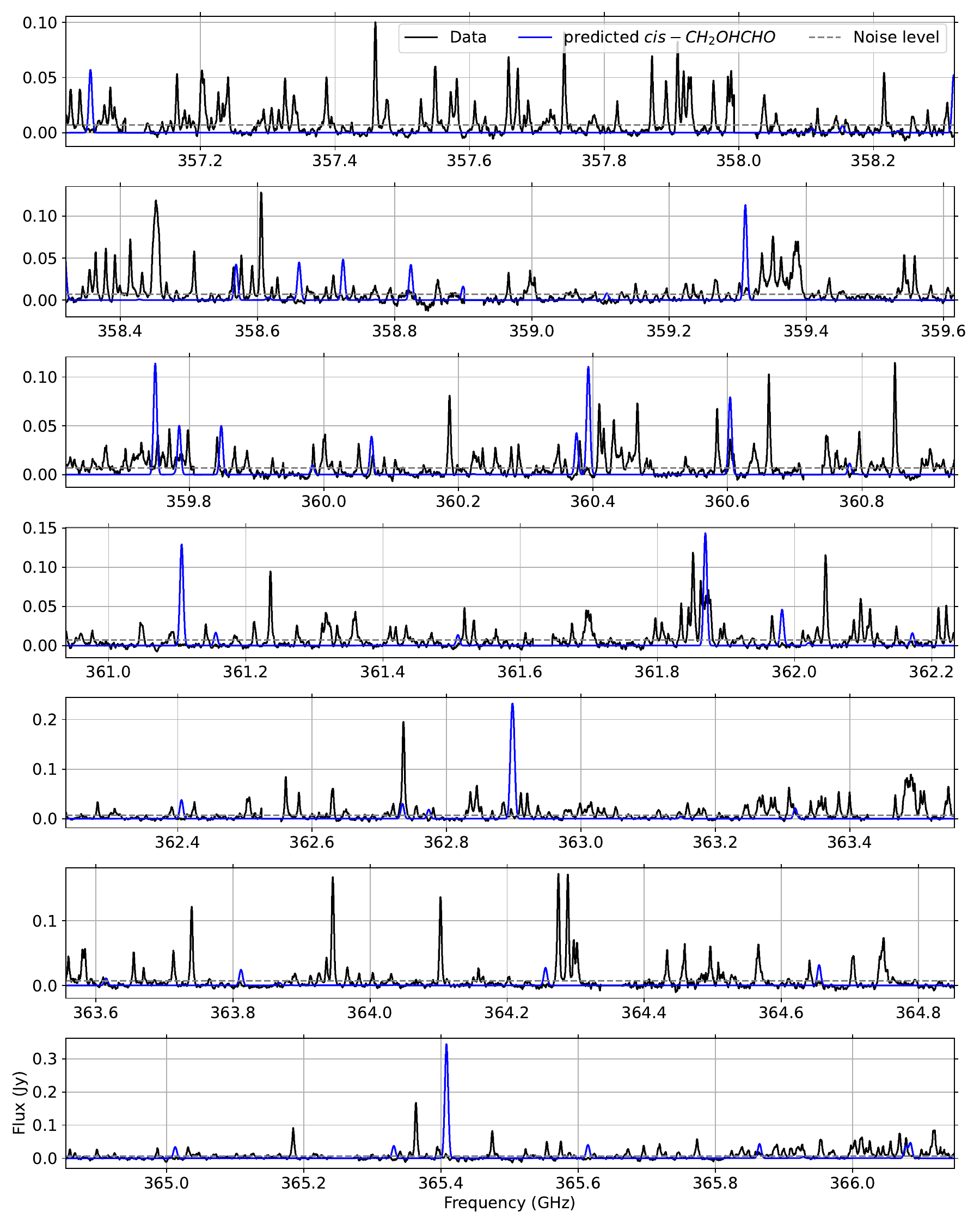}{0.87\textwidth}{}}
\caption{Continuation of figure \ref{glad}.\label{glad1}}
\end{figure}

\clearpage


\section{Appendix E}\label{AppendixE}

To test the assumption of a constant FWHM in our model, we measured the line widths (FWHM) of several molecular transitions. The selected transitions are well isolated from any other blended transitions, whether from the same molecule or from other molecules. The FWHM values were obtained by fitting Gaussian profiles to each of the selected transitions using Python’s \textit{scipy.optimize.curve\_fit} before applying kinematic corrections. The results are presented in Table \ref{tab:FWHM}.

\begin{deluxetable*}{cccccccc}[!htbp]
\tablecaption{Measured FWHM values for different molecular lines \label{tab:FWHM}}
\tablewidth{0pt}
\tablehead{
\colhead{Molecule} & 
\colhead{Frequency} & 
\colhead{FWHM} \\ 
\colhead{} & 
\colhead{GHz} & 
\colhead{$\mathrm{km\; s^{-1}}$} } 
\startdata
$\mathrm{HCN}$ & 354.5055 & $3.72\pm0.08$ \\
$\mathrm{HNC}$ & 362.6303 & $2.69\pm 0.25$ \\
$\mathrm{H_2CO}$ & 362.7361 & $3.97\pm 0.18$ \\
$\mathrm{CH_3OH}$ & 363.7399 & $4.54\pm 0.24$ \\
$\mathrm{CH_3CHO}$ & 351.1187 & $5.42\pm 0.55$ \\
$\mathrm{CH_3OCH_3}$ & 358.5844 & $5.39\pm 0.58$\\
\enddata
\end{deluxetable*}




\end{document}